\documentclass[preprint,12pt,number,square,sort&compress]{elsarticle}

\usepackage[legalpaper, margin=1in]{geometry}
\usepackage{natbib}
\usepackage[separate-uncertainty=true,multi-part-units=single]{siunitx}
\usepackage{blindtext}
\usepackage{graphicx}
\usepackage[export]{adjustbox}
\graphicspath{ {Figures/} }
\usepackage{bm}
\usepackage{amsmath}
\usepackage{amssymb}
\usepackage[normalem]{ulem}
\usepackage{epstopdf}
\usepackage{caption, subcaption}
\usepackage{hyperref}
\hypersetup{
	colorlinks=true,
	linkcolor=cyan,
	filecolor=magenta,      
	urlcolor=blue,
}

\begin{document}
	\let\Gamma\varGamma
	
	\newcommand{\rood}[1]{\textcolor{red}{[#1]}} 
	\newcommand{\gre}[1]{\textcolor[rgb]{0,0.35,0}{[#1]}} 
	\newcommand{\revs}[1]{\textcolor{blue}{#1}} 
	\newcommand{\vect}[1]{\bm{#1}}
	\newcommand{\dd}{\mathrm{d}}
	\newcommand{\ii}{\mathrm{i}}
	\newcommand{\ee}{\mathrm{e}}
	\newcommand{\area}{\mathcal{A}}
	\newcommand{\vol}{\mathcal{V}}
	\newcommand{\saa}{\alpha}
	\newcommand{\sbb}{\beta}
	\newcommand{\dgr}{\dagger}
	\newcommand{\eps}{\varepsilon}
	\newcommand{\sig}{\sigma}
	\newcommand{\kap}{\kappa}
	\newcommand{\kapt}{\tilde{\kappa}}
	\newcommand{\St}{\tilde{S}}
	\newcommand{\gam}{\gamma}
	\newcommand{\Gam}{\Gamma}
	\newcommand{\om}{\omega}
	\newcommand{\TM}{\mathsf{T}}
	\newcommand{\SM}{\mathsf{S}}
	\newcommand{\PM}{\mathsf{M}}
	\newcommand{\beginsupplement}{%
		\setcounter{table}{0}
		\setcounter{figure}{0}
		\setcounter{section}{0}
		\setcounter{subsection}{0}
		\setcounter{equation}{0}
		\setcounter{paragraph}{0}
		\setcounter{page}{1}
		\renewcommand{\thetable}{S\arabic{table}}%
		\renewcommand{\thefigure}{S\arabic{figure}}%
		\renewcommand{\thesection}{S\arabic{section}}%
		\renewcommand{\thesubsection}{S\arabic{section}.\arabic{subsection}}%
		\renewcommand{\theequation}{S\arabic{equation}}%
		\renewcommand{\theparagraph}{S\arabic{paragraph}}%
		\renewcommand{\thepage}{S\arabic{page}}%
		\renewcommand{\theHtable}{S\the\value{table}}%
		\renewcommand{\theHfigure}{S\the\value{figure}}%
		\renewcommand{\theHsection}{S\the\value{section}}%
		\renewcommand{\theHsubsection}{S\the\value{subsection}}%
		\renewcommand{\theHequation}{S\the\value{equation}}%
		\renewcommand{\theHparagraph}{S\the\value{paragraph}}%
	}
	\begin{frontmatter}
		
		\title{Overall dynamic properties of locally resonant viscoelastic layered media  based on consistent field integration for oblique anti-plane shear waves}
		
		\author{Vahidreza Alizadeh and Alireza V. Amirkhizi*}
		\address{Department of Mechanical Engineering\\
			University of Massachusetts, Lowell\\
			One University Avenue, Lowell, MA 01854\\
			*Corresponding author: \href{mailto:alireza\_amirkhizi@uml.edu}{alireza\_amirkhizi@uml.edu}}
		\date{\today}
		
		\begin{abstract}
			
			In this work, the analysis of oblique anti-plane shear waves propagation and scattering in low frequency resonant micro-structured layered media with viscoelastic constituent layers is presented. The band structure of the infinitely periodic systems and scattering off a finite thickness slab of such media are determined using the transfer matrix method. A consistent dynamic field homogenization approach is applied, in which the micro-scale field equations are integrated and the overall macro-scale quantities are defined to be compatible with these integral forms. A reduced set of constitutive tensors is presented for general asymmetric repeating unit cells (RUC), utilizing the proposed homogenized macro-scale quantities combined with Onsager's principle and presumed material form of elastodynamic reciprocity. This set can be further restricted by studying the form of the dispersion equation leading to a unique constitutive tensor. It is shown that for an asymmetric RUC, the full constitutive tensor, including off-diagonal elastic moduli and Willis coupling terms are required in order to match the scattering and band structure of the micro-structured media, but all the off-diagonal parameters vanish for a symmetric RUC. Therefore, it is possible to create an equivalent homogenized representation with a uniquely determined diagonal constitutive tensor for a symmetric RUC, though, as is the case also with asymmetric RUCs, all non-zero components will be wave-vector dependent. Numerical examples are presented to demonstrate the application and consistency of the proposed method. All the diagonal terms are converging to their appropriate Voigt or Reuss averages at the long-wavelength limit for all wave directions. The wave-vector dependent nature of the off-diagonal coupling constants can still be observed even in this limit. The conditions for lossy (passive) or lossless (fully elastic) systems are presented and are shown to impose weak requirements on overall constitutive tensors. 
		\end{abstract}
	\end{frontmatter}
	
	\section{Introduction}
	Static homogenization techniques such as dilute distribution or self-consistent method can be applied in wave propagation problems to determine the overall compliance/stiffness of a medium when the wavelength of interest is much larger than the micro-structural length scale,  \cite{nemat1999micromechanics}. However, at very low wavelengths solids must be viewed as periodic arrays of atoms rather than a homogeneous and continuous medium \cite{kittel2004introduction}. In this regime, the heterogeneity cannot be averaged out using homogenization techniques and one cannot assign overall properties to the medium. Therefore, dynamic homogenization techniques are applicable only for wavelengths below such frequencies. Furthermore, while the apparent overall properties are naturally frequency-dispersive \cite{AnkitSrivastava2015}, the limiting behavior for small frequencies should approach static homogenization limits. Moreover, any overall constitutive description must  be compatible with the observed scattering response of a specimen, i.e., a homogeneous specimen of the same geometry with the calculated overall parameters should produce exactly the same scattering. 
	
	Strong frequency-dispersion is observed in periodic composites and  metamaterials \cite{S.Nemat-Nasser1972, Sigalas1993, Kushwaha1993, Potel1995, Lu2009, Zhu2014, Nassar2016, Zangeneh-Nejad2019}. The  wave propagation in periodic systems is generally analyzed using Bloch formulation and a variety of different approaches have been employed to derive the overall frequency-dispersive constitutive properties of such composites. Dynamic homogenization using two-scale asymptotic methods combined with the Floquet-Bloch approach was proposed in a variety of studies \cite{allaire1998bloch, cherednichenko2007bloch, craster2010high, comi2020homogenization}. In the two-scale asymptotic methods, the macro-scale and local fields are formulated by ``slow" and ``fast" variables, respectively \cite{vkuznetsov2007scattering}. Meng and Guzina \cite{meng2018dynamic} compared such framework with Willis effective description and stated that in the absence of body forces, the two approaches match in long-wavelength low-frequency regime. Higher order asymptotic analyses are also beneficial when wavelength of interest is comparable with the size of heterogeneities \cite{andrianov2008higher}. In scattering-based parameter retrieval methods, the properties of a layered medium are obtained by comparing the complex reflection and transmission coefficients off a finite slab with those from equivalent homogeneous medium \cite{smith2005electromagnetic, Fokin2007,Alu2011,Bayatpur2012,Amirkhizi2017,Amirkhizi2018,Abedi2020}. 
	Although, retrieval methods are commonplace due to their simplicity, they should be carefully applied to ensure satisfaction of causality and passivity requirements \cite{smith2002determination, Chen2004, Simovski2009}. 
	An example of this for long-wavelength limit homogenization of acoustic metamaterials was shown by Li et al. \cite{Li2004}, who utilized coherent potential approximation (CPA) method to estimate the effective density and bulk modulus. They showed theoretical double negative frequency bands in a lossless composite using the CPA approach, in which the heterogeneous system is assumed to be embedded within an effective medium whose material properties are the desired quantities to be extracted, and the scattering is set to zero in the lowest order of frequency. Other authors also used similar and improved techniques to come up with the effective properties in acoustic and electromagnetic metamaterials beyond the long-wavelength limit \cite{Wu2006, Wu2007, Hu2008, Torrent2011, Yang2014}.
	
	In field averaging techniques, wave equations of periodic elastic composites can be explicitly integrated. The objective of this process is to determine integrated, overall, field quantities, that would satisfy wave equations for a presumed homogenized material, and therefore to derive homogenized constitutive properties. It is a natural requirement to expect that the scattering of a fictitious specimen with these homogenized constitutive properties to match those of a micro-structured specimen of the same overall dimensions and geometry. For application of this approach, e.g., using line and surface integrals in electromagnetics see \cite{Smith2006, Amirkhizi2008}, using volume averages in \cite{Amirkhizi2008_CRM}, and using volume averages in elastodynamics see \cite{Willis2009, Nemat-Nasser2011b, Amirkhizi2017}. It is customary to check the band structure or dispersion curves matching between the micro-structured and homogenized media, which is a necessary condition for the matching of scattering response, but not a sufficient one; see \cite{Amirkhizi2017}. Whether field averaging or purely scattering-based techniques are used, it is often needed to describe the overall constitutive properties in the ``Willis-form'' \cite{willis1981variational, Willis1983, willis1984variational, Willis2009, Willis2011}. Willis-type constitutive tensor formalism is a generalization of classical elastic constitutive tensor. In one popular choice of formulation, stress within the homogenized material is coupled to particle velocity and momentum density is coupled to strain through:
	\begin{equation}
		\begin{aligned}
			p_i &= \rho_{ij}v_j + \St_{ijk}\eps_{jk},\\
			\sigma_{ij} &= S_{ijk}v_k + C_{ijkl}\eps_{kl}.
		\end{aligned}
	\end{equation}
	An alternative formulation is used here so that the quantities that have to satisfy continuity requirements (stress and particle velocity) are grouped together:
	\begin{equation}
		\begin{aligned}
			v_i &= \eta_{ij}p_j + \kapt_{ijk}\eps_{jk},\\
			\sigma_{ij} &= \kap_{ijk}p_k + \mu_{ijkl}\eps_{kl}.
		\end{aligned}
	\end{equation}
	The two formulations can be converted to each other via 
	\begin{align*}
		\eta_{ij} \rho_{jk} &= \delta_{ik},   \\
		\kapt_{ijk} &= -\eta_{in}\St_{njk},    \\
		\kap_{ijk} &= S_{ijn}\eta_{nk},\\
		\mu_{ijkl} &= C_{ijkl} - S_{ijm}\eta_{mn}\St_{nkl}.
	\end{align*}
	Srivastava and Nemat-Nasser \cite{Srivastava2012} also use the same grouping adopted here but the dependent and independent variables are switched in that reference. 
	This kind of coupling is analogous to bianisotropy in electromagnetic metamaterials \cite{Sieck2017}. Willis-type coupling has recently been experimentally studied in both 1D \cite{koo2016acoustic, muhlestein2017experimental, Melnikov2019} and 2D metamaterial structures \cite{li2018systematic}. 
	
	This work builds on our previous study \cite{Amirkhizi2018} where we utilized the transfer matrix method (TMM) to calculate exact complex scattering coefficients for oblique incidence of anti-plane shear (shear horizontal or SH) waves off a symmetric layered medium. Versions of TMM has also been successfully applied on half-space stratified media  consisting of a small and large number of anisotropic layers with horizontally polarized surface waves \cite{kuznetsov2004love, kuznetsov2006love}. It was established that any overall constitutive description for such a system will have to be non-local (spatially dispersive) even at the long-wavelength limit (non-zero $f \to 0$). With this requirement, there are infinitely many overall representations that would reproduce the scattering response. In particular, a spatially dispersive non-coupled description will be enough for symmetric systems. Here, we explore application of field integration technique to such problems. Such an approach may require less computational effort in extraction process, particularly when expanded beyond layered systems. 3-phase layered unit cells are  studied here as their band structure and scattering will include low frequency local resonances (in contrast with mostly Bragg bands in 2-phase example). The present study also includes asymmetric cells which fully utilized the integration scheme. The integration scheme is applied consistently on the wave equations to produce the homogenized properties of a layered structure.
	The integration of micro-scale fields provides a consistent description of marco-scale quantities that are not directly accessible from scattering measurements. Determination of such macro-scale field values is used to reduce the dimensionality of the set of possible constitutive tensors, were they to be calculated based on scattering. We further consider the restrictions due to Onsager reciprocity and show that they reduce the space of possible constitutive tensors. While Betti-Maxwell reciprocity is not fully applicable to non-local material descriptions, we formally use them to further restrict the set of possible constitutive tensors to a 1D class. Finally, we reduce this to a single constitutive tensor by excluding the possibility of having multiple solutions for the same value of wave vector conserved component. The possibility of appearance of second solution is discussed in depth in \cite{agranovich_crystal_1984} in electromagnetics of spatially dispersive media. The single solution thus obtained reproduced the scattering response of the micro-structured medium perfectly.  
	The unique constitutive tensor also becomes diagonal for symmetric cells, and coupling terms vanish, as also earlier shown in 1D longitudinal systems by Amirkhizi \cite{Amirkhizi2017}.
	
	\section{Oblique anti-plane shear waves in laminated and Willis-type media}
\begin{figure}[!ht]
		\centering\includegraphics[height=130pt,center]{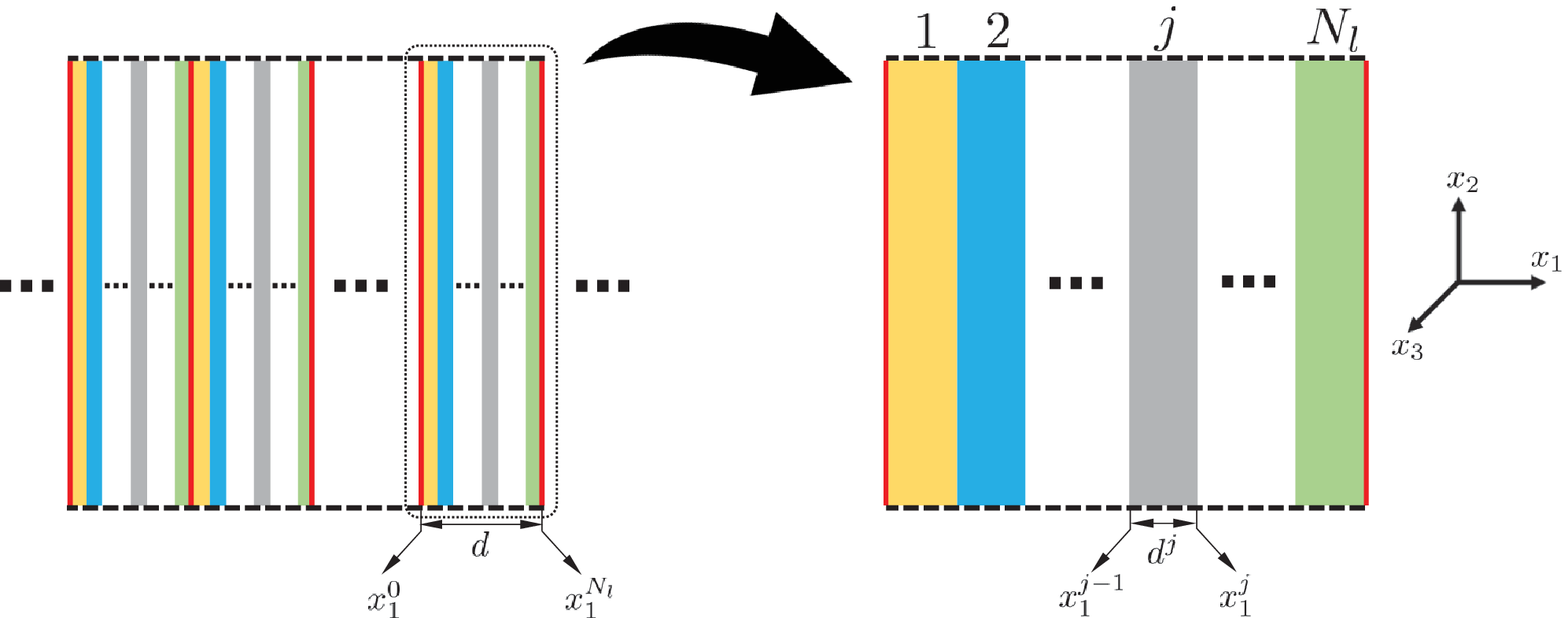}
		\caption{\label{fig:mat_simp} A layered medium with an infinite number of unit cells ($N_c\rightarrow\infty$), where $d$ is the total thickness of each unit cell, is shown here. Each unit cell consists of $N_l$ layers, where $d^j$ represents the thickness of layer $j$.}
	\end{figure}

	The focus of this paper is the study of anti-plane shear wave propagation in layered media. The medium of interest consists of a finite or infinite array of repeating unit cells, each consisting of a number of 2D infinite slabs of different homogeneous materials (parallel to $x_2 x_3$ coordinate plane) stacked along the $x_1$ direction. Anti-plane shear waves with particle velocity polarization in the $x_3$ directions are considered as they propagate in the $x_1 x_2$ plane. It is assumed that all constituent materials have high enough symmetry to ensure the pure (decoupled) anti-plane shear modes exists, e.g., each constituent layer is isotropic. Within a repeating unit cell (RUC), each layer, $1 \leq j \leq N_l$, has thickness $d^j = x_1^j - x_1^{j-1}$, where $x_1^{j-1}$ and $x_1^j$ indicate the interface coordinates of the layer and $N_l$ is the number of layers in one RUC. With respect to $x_3$ direction, this may be considered a decoupled shear horizontal (SH) wave (see Fig.~\ref{fig:mat_simp}). 
	
	\paragraph{Notation}
	The particle velocity, $v_3$, and shear stress, $\sigma_{13}$, are continuous across all interfaces. For a wave vector $\vect{k} = k_1 \vect{e}_1 + k_2 \vect{e}_2$ we introduce the notation
	\begin{equation}
		\vect{\psi}^j(x_1, x_2, k_2) =
		\begin{pmatrix}
			v_3\\
			\tau_5
		\end{pmatrix}
		= \vect{\zeta}^j(k_2)\vect{\delta}^j(x_1,k_2)\vect{A}^j \ee^{-\ii(k_2 x_2)},
	\end{equation}
	as the state vector representing the continuous quantities in layer $j$. Here we have used Voigt notation for stress, $\tau_5 = \sigma_{31} = \sigma_{13}$. The equations are written for harmonic steady state oscillation with angular frequency $\omega$ and physical quantities are the real part of the complex phasor form with $\ee^{\ii(\omega t-\vect{k.x})}$.  $\ee^{\ii\omega t}$ is generally dropped in all equations when not necessary (thought it clearly controls the time derivatives) but the spatial part is explicitly handled in most cases. The polarization and phase advance matrices are defined for parameter $k_2$ as 
	\begin{equation}
		\begin{aligned}
			\vect{\zeta}^j(k_2) &= 
			\begin{pmatrix}
				1 & 1\\
				-Z^{j+} & -Z^{j-}
			\end{pmatrix}, \\ 
			\vect{\delta}^j(x_1,k_2) &=
			\begin{pmatrix}
				\ee^{-\ii k_1^{j+}(x_1 - x_1^{j-1})} & 0\\
				0 & \ee^{-\ii k_1^{j-}(x_1 - x_1^{j-1})}
			\end{pmatrix},
		\end{aligned}
	\end{equation}
	where superscripts $^+$ and $^-$ identify the two anti-plane shear waves traveling presumably towards the positive and negative $x_1$ values (in the sense of phase advance or power flux). To satisfy the continuity for all values of $x_2$, $k_2$ has to be a fixed parameter over the entire domain. But $k_1$ takes different values for each constituent material $k_1^{j\pm} = \pm\sqrt{\left(\omega/c_T^j\right)^2 - k_2^2}$, in which $c_T^j$ is the shear wave speed of layer $j$. 
	Of course, for reciprocal constituent materials one can write $k_1^{j-} = - k_1^{j+} = -k_1^j$. $\vect{A}^j = (A^{j+}, A^{j-})^{\top}$ is the amplitude vector of the two traveling waves and $Z^{j\pm} = \mu_{55}^j k_1^{j\pm}/\omega = \pm\mu_{55}^j\sqrt{\left(1/c_T^j\right)^2 - s_2^2}$ are the impedances, where $s_2 = k_2/\omega$ is the slowness vector component along $x_2$-axis and the Voigt notation is again used for the elastic moduli, i.e. $\mu_{55} = C_{1313}$ in standard tensorial notation. 
	\paragraph{Constitutive and dispersion relations for oblique anti-plane shear waves in Willis-type media}
	The Willis-type constitutive law for oblique incident anti-plane shear horizontal (SH) waves can be written as \cite{Amirkhizi2018}, 
	\begin{equation}\label{eq:will}
		\begin{aligned}
			V_i &= \eta_{ij}P_j + \kapt_{ijk}E_{jk},\\
			\Sigma_{ij} &= \kap_{ijk}P_k + \mu_{ijkl} E_{kl},
		\end{aligned}
	\end{equation}
	where capital Roman and Greek letters are used to implicitly represent the presumed apparent or macro-scale quantities in a micro-structured medium. $\Sigma_{ij}$, $E_{ij}$, $V_j$ and $P_j$ represent stress, strain, particle velocity and momentum density components, respectively. $\eta_{ij}$ represents the appropriate specific volume (inverse mass density), $\mu_{ij}$ are the moduli of elasticity, while $\kapt_{ijk}$ denotes particle velocity/strain and $\kap_{ijk}$ represents stress/momentum density couplings. The similar small letters will represent local field quantities. The couple stresses are assumed negligible and similarly their conjugate rotations are not included in this analysis, therefore the stress and strain tensors are considered symmetric and in later parts of this paper the Voigt notation will be adopted. 
	Specifically, with the focus on oblique anti-plane shear waves the expansion of Eq.~\eqref{eq:will}, yields,
	\begin{equation}\label{eq:const}
		\begin{pmatrix}
			V_3 \\
			T_4 \\
			T_5
		\end{pmatrix}
		=
		\begin{pmatrix}
			\eta_{33} & \kap_{34} & \kap_{35}\\ 
			\kap_{43} & \mu_{44} & \mu_{45} \\
			\kap_{53} & \mu_{54} & \mu_{55}
		\end{pmatrix}
		\begin{pmatrix}
			P_3\\ 
			\Gam_4 \\
			\Gam_5
		\end{pmatrix},
	\end{equation}
	where, Voigt notation is used for strain $\Gam_4 = 2 E_{23} = 2 E_{32},~\Gam_5 = 2 E_{31} = 2 E_{13}$, and shear stress $T_4 = \Sigma_{23} = \Sigma_{32},~T_5 = \Sigma_{31} = \Sigma_{13}$ components. Superscript $\tilde{}$ is dropped for the sake of simplicity as the subscripts uniquely identify them. Therefore, the dispersion relation for oblique SH waves in any material that can be described by a Willis-type linear constitutive law and that allows for anti-plane shear waves can be derived in terms of slowness vector components ($\vect{s} = \vect{k}/\omega$) as \cite{Amirkhizi2018},
	\begin{equation}\label{eq:disp}
		\begin{split}
			0 = 1 &+ (\kap_{35}+\kap_{53})S_1 + (\kap_{34}+\kap_{43})s_2
			+ \left[\kap_{34}\kap_{53}+\kap_{43}\kap_{35}-\eta_{33}(\mu_{45}+\mu_{54})\right]S_1 s_2 \\
			&+(\kap_{35} \kap_{53} - \eta_{33}\mu_{55}) S_1^2 +(\kap_{34} \kap_{43} - \eta_{33}\mu_{44}) s_2^2. 
		\end{split}
	\end{equation}
	The component $s_2$ is constant over the entire domain, but $S_1$ represents an apparent overall quantity as $s_1^j$ is different in each material layer. All components of this tensor are, in general, functions of the frequency and wave vector.
	
	\section{Homogenized Willis constitutive tensor extraction using field integration}
	The local conservation and kinematic equations for complex amplitudes of the fields in oblique anti-plane shear waves are
	\begin{equation}
		\begin{aligned}
			v_3 &= \ii\omega u_3,\\
			\gamma_4 &= \dfrac{\partial u_3}{\partial x_2} = -\ii k_2u_3 = -\dfrac{k_2}{\omega}v_3,\\
			\gamma_5 &= \dfrac{\partial u_3}{\partial x_1} = \dfrac{1}{\ii\omega}\dfrac{\partial v_3}{\partial x_1},\\
			\dfrac{\partial \tau_5}{\partial x_1} + \dfrac{\partial \tau_4}{\partial x_2} &= \ii\omega p_3 \Rightarrow \dfrac{1}{\ii\omega}\dfrac{\partial \tau_5}{\partial x_1} - \dfrac{k_2}{\omega}\tau_4 = p_3.
		\end{aligned}
	\end{equation}
	All these quantities are locally functions of $x_1$ but are also parameterized by $k_2$, and $\omega$.  We also define the slowness component in $x_2$ direction $s_2 = k_2 / \omega$.\\
	In source-free harmonic analysis, any oblique anti-plane shear wave traveling in a layered medium can be written as superposition of two linearly independent waves. We choose for this basis the eigenvectors of the transfer matrix when a finite or infinite structure consists of a periodic array of a repeating unit cell. Such eigenvectors generally satisfy the Bloch-Floquet periodicity
	\begin{equation}
		\mathcal{G}^\alpha(x_1^n) = \mathcal{G}^\alpha(x_1^0)\ee^{-\ii Q^\alpha},
	\end{equation}
	where, $\mathcal{G}$ represents any field quantity, namely displacement ($u_3$), particle velocity ($v_3$), strain component ($\gamma_4,\gamma_5$), stress component ($\tau_4,\tau_5$), or momentum ($p_3$), and the quantity $Q^\alpha$ can be calculated from the components of the transfer matrix by eigenvalue analysis and its real part is normally referred to as phase advance across an RUC (for a more detailed discussion on the transfer matrix method, see \ref{app:tm}). The superscript $\alpha$ represents the two independent solutions. In the following we have used the convention $\alpha = \pm$ to represent positive and negative $x_1$ direction of the eigenvector waves in the sense of phase propagation or power flux.
	For a finite specimen consisting of $N_c$ unit cells with boundaries at $x_1^0$ and $x_1^0 + N_c d$, it is desirable to match the continuous micro-structural quantities with yet-to-be defined macro-scale apparent quantities on these surfaces ($x_1 = const.)$. Such a requirement maintains the nature of continuity requirements at the boundaries with homogeneous semi-infinite domains, which control the scattering off of finite specimens. If other schemes are used instead, the continuity boundary conditions will have different form and nature for the macro-scale quantities in the homogenized system compared to those in the micro-structured system. Therefore,
	\begin{equation}
		\begin{aligned}
			V_3^\alpha(x_1^0) &= v_3^\alpha(x_1^0),\\
			T_5^\alpha(x_1^0) &= \tau_5^\alpha(x_1^0),
		\end{aligned}
	\end{equation}
	and similarly for $x_1^0 + N_c d$. Assuming that the apparent response may be considered simply as the propagation of a source free harmonic wave in a homogeneous medium, these macro-scale fields can therefore be written everywhere as
	\begin{equation}
		\begin{aligned}
			V_3^\alpha(x_1) &= v_3^\alpha(x_1^0) \ee^{\ii\phi^\alpha},\\
			T_5^\alpha(x_1) &= \tau_5^\alpha(x_1^0)\ee^{\ii\phi^\alpha},
		\end{aligned}
	\end{equation}
	where, $\phi^\alpha = - K_1^\alpha (x_1 - x_1^0) - k_2x_2$, with $K_1^\alpha d = Q^\alpha \pm 2 n \pi$. Therefore, as it is also pointed out in Eq.~\eqref{eq:tm-Z}, the impedance is $Z_{53}^\alpha = -T_5^\alpha/V_3^\alpha$.
	Furthermore, the following also need to be satisfied for such an apparent macro-scale wave,
	\begin{equation}\label{eq:heq}
		\begin{aligned}
			\Gamma_4^\alpha(x_1) &= -\dfrac{k_2}{\omega}V_3^\alpha,\\
			\Gamma_5^\alpha(x_1) &= \dfrac{1}{\ii\omega}\dfrac{\partial V_3^\alpha}{\partial x_1} = -\dfrac{K_1^\alpha}{\omega}V_3^\alpha,\\
			\dfrac{\partial T_5}{\partial x_1} + \dfrac{\partial T_4}{\partial x_2} &= \ii\omega P_3^\alpha \Rightarrow \dfrac{K_1^\alpha}{\omega}T_5^\alpha + \dfrac{k_2}{\omega}T_4^\alpha = -P_3^\alpha.
		\end{aligned}
	\end{equation}
	$\Gamma_4^\alpha(x_1)$ requirement matches well with microscopic consideration $\gamma_4 = -s_2 v_3$, and therefore the same approach as with the continuous quantities will be used for $\Gamma_4$, i.e., define
	\begin{equation}
		\Gamma_4^\alpha(x_1) = \gamma_4^\alpha(x_1^0) \ee^{\ii\phi^\alpha}.
	\end{equation}
	To satisfy the other two conditions consider integration of the microscopic quantities over a unit cell,
	\begin{equation*}
		\begin{aligned}
			\displaystyle\int_{\Omega}\gamma_5^\alpha \dd x_1 = \dfrac{1}{\ii\omega}\displaystyle\int_{\Omega}\dfrac{\partial v_3^\alpha}{\partial x_1} \dd x_1
			&= \dfrac{1}{\ii\omega}\left[v_3^\alpha(x_1^0 + d) - v_3^\alpha(x_1^0)\right]
			= \dfrac{1}{\ii\omega}\left[V_3^\alpha(x_1^0 + d) - V_3^\alpha(x_1^0)\right]\\
			&= \dfrac{1}{\ii\omega}\left(\ee^{-\ii K_1^\alpha d} - 1\right)V_3^\alpha(x_1^0)\\
			&= -\dfrac{Q^\alpha}{\omega}\ee^{-\ii Q^\alpha /2}\left(\dfrac{\ee^{-\ii Q^\alpha /2} - \ee^{\ii Q^\alpha /2}}{2\ii}\right)\left(\dfrac{1}{-Q^\alpha /2}\right)V_3^\alpha(x_1^0)\\
			&= -\dfrac{Q^\alpha }{\omega}\chi(Q^\alpha /2)\ee^{-\ii Q^\alpha /2}V_3^\alpha(x_1^0),
		\end{aligned}
	\end{equation*}
	where, $\Omega = \{x_1: x_1^0 \leq x_1\leq x_1^{N_l} = x_1^0 + d\}$, with $N_l$ and $d$ being the number of layers and the total size of the unit cell, respectively, and $\chi(\beta) = \sin(\beta)/\beta$. To satisfy Eq.~\eqref{eq:heq}, one can set the apparent macro-scale strain
	\begin{equation}
		\Gamma_5^\alpha(x_1^0) = \dfrac{\dfrac{1}{d}\displaystyle\int_{\Omega}\gamma_5^\alpha \dd x_1}{\chi(Q^\alpha /2)}\ee^{\ii Q^\alpha /2}.
	\end{equation}
	For the macro-scale wave $\Gamma_5^\alpha(x_1)$ must be related by a simple phase difference to $\Gamma_5^\alpha(\bar{x}_1)$, where $\bar{x}_1 = x_1^0 + (d/2)$. At this point,
	\begin{equation}\label{eq:Gam5Int}
		\Gamma_5^\alpha(\bar{x}_1) = \Gamma_5^\alpha(x_1^0)\ee^{-\ii Q^\alpha /2} = \dfrac{\dfrac{1}{d}\displaystyle\int_{\Omega}\gamma_5^\alpha \dd x_1}{\chi(Q^\alpha /2)}.
	\end{equation}
	Note that one can similarly set
	\begin{equation}
		\begin{aligned}
			P_3^\alpha(\bar{x}_1) &= \dfrac{\dfrac{1}{d}\displaystyle\int_{\Omega}p_3^\alpha \dd x_1}{\chi(Q^\alpha /2)},\\
			T_4^\alpha(\bar{x}_1) &= \dfrac{\dfrac{1}{d}\displaystyle\int_{\Omega}\tau_4^\alpha \dd x_1}{\chi(Q^\alpha /2)}.
		\end{aligned}
	\end{equation}
	On the other hand, it is physically more desirable to ensure the integral of the macro-scale momentum $P_3^\alpha(x_1)$ in any unit cell matches the total micro-structural quantity. This is satisfied by the previous definition since
	\begin{equation*}
		\begin{aligned}
			\dfrac{1}{d}\displaystyle\int_{\Omega}p_3^\alpha \dd x_1 &= \dfrac{1}{d}\int_{\Omega} P_3^\alpha(\bar{x}_1)\ee^{-\ii K_1^\alpha(x_1 - \bar{x}_1)} \dd x_1
			= P_3^\alpha(\bar{x}_1)\int_{-d/2}^{d/2} \ee^{-\ii K_1^\alpha\xi} \dd\xi\\
			&= P_3^\alpha(\bar{x}_1)\left(\dfrac{\ee^{-\ii Q^\alpha /2} - \ee^{\ii Q^\alpha /2}}{-2\ii Q^\alpha /2}\right)
			= P_3^\alpha(\bar{x}_1)\chi(Q^\alpha /2).
		\end{aligned}
	\end{equation*}
	Similarly, the proposed macro-scale definitions for $\Gamma_5$ and $T_4$ are physically satisfactory as in both cases the integration along the $x_1$ direction across the unit cell can be related to physical quantities, such as displacement differential and total traction force on the relevant planes. 
	
	The constitutive description of each isotropic layer can be summarized as
	\begin{equation*}
		\begin{pmatrix}
			v_3\\
			\tau_4\\
			\tau_5
		\end{pmatrix}^j=
		\begin{pmatrix}
			\eta_{33} & 0 & 0\\
			0 & \mu_{44} & 0\\
			0 & 0 & \mu_{55}
		\end{pmatrix}^j
		\begin{pmatrix}
			p_3\\
			\gamma_4\\
			\gamma_5
		\end{pmatrix}^j,
	\end{equation*}
	where $\eta_{33} =  1/\rho$ represents the specific volume and $\mu_{44} = \mu_{55}$ is the shear modulus. The macroscopic quantities that are not directly part of the eigen-vectors of the transfer matrix can be written as,
	\begin{equation*}
		\begin{pmatrix}
			\gamma_5\\
			p_3\\
			\tau_4
		\end{pmatrix}^j = \vect{\xi}^j
		\begin{pmatrix}
			v_3\\
			\tau_5
		\end{pmatrix}^j; \quad \vect{\xi}^j =
		\begin{pmatrix}
			0 & \mu_{55}^{-1}\\
			\eta_{33}^{-1} & 0\\
			-\mu_{44}s_2 & 0
		\end{pmatrix}^j.
	\end{equation*}
	Collectively and over a layered repeating unit cell the macro-scale quantities can be written as
	\begin{equation}\label{eq:int_fin}
		\begin{pmatrix}
			\Gamma_5\\
			P_3\\
			T_4
		\end{pmatrix}^\alpha(\bar{x}_1) = 
		\dfrac{1}{\chi(Q^\alpha /2)}\dfrac{1}{d}\sum_{j = 1}^{N_l}\vect{\xi}^j\vect{\zeta}^j\left[\int_{x_1^{j-1}}^{x_1^j}\vect{\delta}^j(x_1,k_2)\dd x_1\right]({\vect{\zeta}^j})^{-1}\vect{\psi}^j(x_1^{j-1},k_2).
	\end{equation}
	Using $\vect{\psi}^j = \vect{\zeta}^j\vect{\delta}^j(d^j,k_2)({\vect{\zeta}^j})^{-1} \vect{\psi}^{j-1}$ for $1 \leq j \leq N_l$ and combined with 
	\begin{equation}
		\begin{pmatrix}
			V_3\\
			\Gamma_4\\
			T_5
		\end{pmatrix}^\alpha(\bar{x}_1) = \ee^{-\ii Q^\alpha /2}
		\begin{pmatrix}
			1 & 0 \\
			-s_2 & 0 \\
			0 & 1
		\end{pmatrix}
		\vect{\psi}^0,
	\end{equation}
	all overall macro-scale field quantities may be calculated at the center of the unit cell, consistent with the equations of motions. 
	
	Equation~\eqref{eq:const} may be considered as a basis to determine the constitutive tensors if all field quantities are known. However, for a single field solution, there are only 3 equations for 9 unknowns. If the system is spatially non-dispersive, one may combine infinitely many solutions for different wave vectors and determine the constitutive tensor components as functions of frequency. Of course, it has been shown that for micro-structured media, this is impossible (except in limited situations) as the infinitely many solutions are not compatible with a single set of constitutive constants, and the system has to be described with spatially dispersive constitutive parameters \cite{Amirkhizi2018}. In the following, we will first utilize the Onsager principle to reduce the level of indeterminacy. Next, we will consider a sufficient form of the Betti-Maxwell reciprocity (which in fact is only applicable to spatially non-dispersive constitutive descriptions) to restrict possible scenarios for the overall properties further. Finally, we will look at the dispersion relation form and propose and determine reasonable restrictions that allow us to solve for a unique tensor of all the homogenized Willis constitutive parameters based on the integrated field quantities and one that exactly reproduce the band structure and scattering of a micro-structured specimen.
	
	\subsection{Onsager Reciprocal Relations}
	It is possible to utilize an extension of Onsager's reciprocity principle for spatially dispersive media. In such cases, the constitutive tensors for wave vectors $\vect{k}$ and $-\vect{k}$ are related to each other \cite{agranovich_crystal_1984, deGroot1984}. Utilizing the Onsager's principle in our case will increase the number of available equations to 6 while keeping the unknowns at 9. Here superscripts $^+$, and $^-$ denote the constitutive parameters for wave vectors $\vect{k}$, and $-\vect{k}$, respectively. Onsager's principle states:
	\begin{equation*}
		\eta_{33}^- = \eta_{33}^+,\qquad \mu_{44}^- = \mu_{44}^+,\qquad  \mu_{55}^- = \mu_{55}^+,
	\end{equation*}
	\begin{equation}\label{eq:onsag}
		\begin{aligned}
			\mu_{54}^- = \mu_{45}^+,&\qquad  \mu_{45}^- = \mu_{54}^+,\\
			\kap_{43}^- = -\kap_{34}^+,&\qquad  \kap_{34}^- = -\kap_{43}^+,\\
			\kap_{53}^- = -\kap_{35}^+,&\qquad  \kap_{35}^- = -\kap_{53}^+. 
		\end{aligned}
	\end{equation}
	Notice the negative sign in the coupling terms. The sign is associated with whether the field quantities will change their sign with time reversal (odd) or not (even), or equivalently whether they depend on micro-scale velocities in and odd or even form. For the constitutive parameters connecting quantities of different parities, the Onsager's principle requires a negative sign \cite{Nassar2020}.
	
	\subsection{Betti-Maxwell Reciprocity}
	Unlike Onsager's reciprocity, the constitutive form of Betti-Maxwell's reciprocity  is really only applicable to spatially non-dispersive media. Nonetheless, it is worth considering its connection to micro-scale material form and relation to macro-scale reciprocity which is applicable for broad classes of system (including all systems consisting of elastic components along with isotropic viscoelastic ones).
	
	To study the micro-scale reciprocity theorem as discussed in \cite{Achenbach2006} for the Willis material described by Eq.~\eqref{eq:will} we write the dynamic equations for two states, $\saa$ and $\sbb$, as follows:
	\begin{equation}
		\begin{aligned}
			\sigma_{ji,j}^\saa + f_i^\saa &= \dot{p}_i^\saa,\\
			\sigma_{ji,j}^\sbb + f_i^\sbb &= \dot{p}_i^\sbb,
		\end{aligned}
	\end{equation}
	where $f_i$ are the components of the body force. The reciprocity theorem can be derived by multiplying the displacement field in one state, e.g., $\saa$, $u_i^\saa$, by dynamic equation of the other state, e.g., $\sbb$, and subtracting it from the equation based on reversing this, \cite{achenbach_2004}:
	\begin{equation}\label{eq:rec}
		\left(\sigma_{ji,j}^\saa u_i^\sbb-\sigma_{ji,j}^\sbb u_i^\saa\right) + \left(f_i^\saa u_i^\sbb-f_i^\sbb u_i^\saa\right) = \dot{p}_i^\saa u_i^\sbb-\dot{p}_i^\sbb u_i^\saa.
	\end{equation}
	The first term can be simplified using the infinitesimal strain formulations as follows:
	\begin{equation*}\label{eq:rec_TU}
		\begin{aligned}
			(\sigma_{ji,j}^\saa u_i^\sbb -\sigma_{ji,j}^\sbb u_i^\saa ) &= \left[\sigma_{ji}^\saa u_i^\sbb -\sigma_{ji}^\sbb u_i^\saa \right]_{,j} - \left(\sigma_{ij}^\saa \eps_{ij}^\sbb -\sigma_{ij}^\sbb \eps_{ij}^\saa \right)\\
			&= \left[\sigma_{ji}^\saa u_i^\sbb -\sigma_{ji}^\sbb u_i^\saa \right]_{,j} -\left(\mu_{ijkl}-\mu_{klij}\right)\eps_{kl}^\saa \eps_{ij}^\sbb -\kap_{ijk}\left(p_k^\saa \eps_{ij}^\sbb -p_k^\sbb \eps_{ij}^\saa \right).
		\end{aligned}
	\end{equation*}
	Furthermore, focusing on time-harmonic cases to swap the time derivative from momentum density to displacement to write the remainder in terms of velocity and using the constitutive equations:
	\begin{equation*}\label{eq:rec_PU}
		\dot{p}_i^\saa u_i^\sbb -\dot{p}_i^\sbb u_i^\saa = p_i^\saa p_k^\sbb \left(\eta_{ik}-\eta_{ki}\right) + \kapt_{ikl}\left(p_i^\saa \eps_{kl}^\sbb - p_i^\sbb \eps_{kl}^\saa\right).
	\end{equation*}
	Thus after simplification Eq.~\eqref{eq:rec} becomes,
	\begin{equation}\label{eq:rec_fin}
		\begin{aligned}
			\left[\sigma_{ji}^\saa u_i^\sbb -\sigma_{ji}^\sbb u_i^\saa \right]_{,j} + \left(f_i^\saa u_i^\sbb-f_i^\sbb u_i^\saa\right) &= \left(\mu_{ijkl}-\mu_{klij}\right)\eps_{kl}^\saa \eps_{ij}^\sbb+ p_i^\saa p_k^\sbb\left(\eta_{ik}-\eta_{ki}\right)\\
			&+ \left(\kap_{ijk}+\kapt_{kij}\right)\left(p_k^\saa \eps_{ij}^\sbb - p_k^\sbb \eps_{ij}^\saa\right).
		\end{aligned}
	\end{equation}
	For most standard materials, the second and third terms on the rhs are identically zero (density or specific volume are scalars and there are no couplings). The first term on the rhs is zero in elastic materials due to the major symmetry as a consequence of the existence of strain energy potential. For viscoelastic dissipative materials, the proof of major symmetry for relaxation modulus and creep compliance demands more sophisticated arguments using the combination of thermodynamics and the application of Onsager's reciprocity relations \cite{DAY1971}, therefore, experimental verification was suggested \cite{Rogers1963}. Of course, certain material symmetries, e.g., isotropy, are stronger (sufficient) conditions  than major symmetry \cite{Rogers1963}. Even without employing the Onsager's reciprocity principle and by utilizing some thermodynamic and analytical restrictions, Matarazzo \cite{Matarazzo2001} showed that major symmetry is guaranteed for a linear viscoelastic solid, if both instantaneous and equilibrium limit tensors have major symmetry. Consequently, for elastic and (broad class of) linear viscoelastic solid constitutive layers with no constitutive coupling and scalar specific volume, the rhs of Eq.~\eqref{eq:rec_fin} would always vanish, which proves that the lhs of Eq.~\eqref{eq:rec_fin} is also zero. This is commonly referred to as the local reciprocity theorem in the literature \cite{Achenbach2006}. The global reciprocity theorem can be derived by integrating the lhs of Eq.~\eqref{eq:rec_fin} over any volume $\vol$,
	\begin{equation}\label{eq:rec_glob}
		\int_{\vol} \left(f_i^\saa u_i^\sbb-f_i^\sbb u_i^\saa\right) \dd\vol + \int_{\mathcal{S}} \left(\sigma_{ji}^\saa u_i^\sbb -\sigma_{ji}^\sbb u_i^\saa \right)n_j \dd\mathcal{S} = 0,
	\end{equation}
	where Gauss's theorem is used and $\mathcal{S} = \partial \mathcal{V}$. Here we have kept the stress form instead of converting, as is normally done, to boundary tractions. In this form, and in contrast with the rhs of the local form, the reciprocity theorem may be purely cast in the form of excitation and response of the system. For that reason, it is reasonable to expect or enforce similar condition when macro-scale fields are defined for the same system:
	\begin{equation}\label{eq:rec_glob_macro}
		\int_{\vol} \left(F_i^\saa U_i^\sbb - F_i^\sbb U_i^\saa\right) \dd\vol + \int_{\mathcal{S}} \left(\Sigma_{ji}^\saa U_i^\sbb - \Sigma_{ji}^\sbb U_i^\saa \right)n_j \dd\mathcal{S} = 0.
	\end{equation}
	The surface integral can be converted to a volume integral. Note that integration domains smaller than homogenization length scale are rather artificial here, but we may proceed with assuming the integrands may be considered identically zero. When macro-scale fields satisfy the relevant equilibrium and compatibility conditions (which is what has been enforced e.g., in our definition of the overall quantities, based on the integration of the field equations), then: 
	\begin{equation} \label{eq:rec-loc-macro}
		\begin{aligned}
			\left[\Sigma_{ji}^\saa U_i^\sbb -\Sigma_{ji}^\sbb U_i^\saa \right]_{,j} + \left(F_i^\saa U_i^\sbb - F_i^\sbb U_i^\saa\right) &= \left(\mu_{ijkl}-\mu_{klij}\right)E_{kl}^\saa E_{ij}^\sbb+ P_i^\saa P_k^\sbb\left(\eta_{ik}-\eta_{ki}\right)\\
			&+ \left(\kap_{ijk}+\kapt_{kij}\right)\left(P_k^\saa E_{ij}^\sbb - P_k^\sbb E_{ij}^\saa\right).
		\end{aligned}
	\end{equation}
	Note that all constitutive parameters here (particularly $\mu$, $\eta$, $\kap$, and $\kapt$) are associated with the macro-scale apparent tensor. To deduce symmetry conditions for these constitutive tensors based on the expectation that this quantity should vanish will require that the strain and momentum density fields to be independently achievable. This is not possible in the absence of body forces (for harmonic waves with no sources). We do not address the question of whether, by prescription of non-zero body forces and boundary conditions, this independence may be accomplished. (Note that for consistency, the macro-scale body force $F_i$ may not have variations that are finer than the macro-scale homogenization limit. In essence, one may approach this by considering unit cell solutions with macro-scale prescribed boundary fields and body forces to be simply harmonic functions of time and space for given frequency and wave vector. But this does not constitute a unique choice.) We note, however, that the symmetry conditions:
	\begin{equation}
		\begin{aligned}
			\eta_{ij} &= \eta_{ji},\\
			\mu_{ijkl} &= \mu_{klij},\\
			\kap_{ijk} &= -\kapt_{kij},
		\end{aligned}
	\end{equation}
	for the macro-scale Willis parameters constitute sufficient conditions to render this quantity zero and therefore will guarantee an apparent (macro-scale) reciprocal system as one might expect. In the context of SH-waves (Eq.~\eqref{eq:const}) and in Voigt notation, this would read as 
	\begin{align}
		\mu_{54} &= \mu_{45},\label{eq:mu_sym}\\
		\kap_{43} &= -\kap_{34},\\
		\kap_{53} &= -\kap_{35}.\label{eq:kap35_anti_sym}
	\end{align}
	Strictly speaking such a requirement is rather limited for spatially dispersive homogenized properties, as unless both $\alpha$ and $\beta$ have the same wave vector $\vect{k}$, this formulation is not applicable. For this reason, the superscripts $\pm$ associated with $\vect{k}$ and $-\vect{k}$ are dropped from this point on to indicate the assumption that the constitutive tensors, while functions of wave vector in general, are the same if one simply reverses the wave vector. This restriction turns out to be fully compatible with both reciprocity restrictions yet not adding any new information or requirements. It is also intuitively compatible with Betti-Maxwell reciprocity. Nevertheless, enforcing the material form of Betti-Maxwell's conditions on the constitutive parameters will lead to well-behaved solutions and reduces the dimensionality of class of possible constitutive tensors. However, while it appears that we now have enough (9) equations to determine all unknowns, the system of equations thus developed is still rank deficient by one.
	\subsection{Uniqueness of the constitutive tensor}
	The combination of the Onsager's principle and material Betti-Maxwell reciprocity as shown in Eq.~\eqref{eq:onsag} and Eqs.~\eqref{eq:mu_sym}--\eqref{eq:kap35_anti_sym} can be used to write: 
	\begin{equation}\label{eq:willis_asym_pm}
		\begin{pmatrix}
			\mathrm{B}^+\\
			\mathrm{B}^-
		\end{pmatrix} = 
		\begin{pmatrix}
			\mathrm{A} & 0\\
			0 & \mathrm{A}
		\end{pmatrix}
		\begin{pmatrix}
			\mathrm{C}^+\\
			\mathrm{C}^-
		\end{pmatrix},
	\end{equation}
	where $\mathrm{B}^\pm = (V_3^\pm, T_4^\pm, T_5^\pm)^\top$,  $\mathrm{C}^\pm = (P_3^\pm, \Gamma_4^\pm, \Gamma_5^\pm)^\top$, and
	\begin{equation}
		\mathrm{A} = 
		\begin{pmatrix}
			\eta_{33}&\kap_{34}&\kap_{35}\\
			-\kap_{34}&\mu_{44}&\mu_{45}\\
			-\kap_{35}&\mu_{45}&\mu_{55}
		\end{pmatrix}.
	\end{equation}
	As it is shown in Eq.~\eqref{eq:willis_asym_pm} both $+$ and $-$ solution sets are defined using the same Willis matrix.  This is a linear system of equation for components of $\mathrm{A}$, but it is always rank deficient by one. Therefore, one more equation is needed to arrive at a unique constitutive tensor. 
	The simplified version of dispersion relation (after taking into account the conditions shown in Eqs.~\eqref{eq:mu_sym}--\eqref{eq:kap35_anti_sym}) becomes
	\begin{equation}
		0 = 1 - 2\left(\kap_{34}\kap_{35} + \eta_{33}\mu_{45}\right)S_1s_2-\left(\kap_{35}^2 + \eta_{33}\mu_{55}\right) S_1^2 -\left(\kap_{34}^2 + \eta_{33}\mu_{44}\right) s_2^2.
	\end{equation}
	As mentioned, while the constitutive coefficients are general functions of slowness or wave vector, it is assumed that a simple reversal of this vector does not affect the constitutive tensor (an intuitive consequence of reciprocity). This is also compatible with the dispersion relation not having any first order terms in terms of the slowness vector components. If the coefficient of $S_1 s_2$ is non-zero, there may exist directions for wave vector (which is the same as that of the slowness vector and can be parameterized as $S_1/s_2$) along which there are potentially two wavelengths can propagate in the system. Such a phenomenon is quite an important indicator of spatial dispersion in electromagnetics \cite{agranovich_crystal_1984}. However, in the present study of layered micro-structured media, there has not been any such observation. Therefore, we consider adding the following equation to our set of equations:
	\begin{equation}\label{eq:s1s2_coeff}
		\kap_{34}\kap_{35} + \eta_{33}\mu_{45} = 0.
	\end{equation}
	It is worth mentioning that in the case of a spatially non-dispersive Willis-matrix the presence of a mirror or $\pi$-rotational symmetry in or around the $x_1$ or $x_2$ directions will also force the coefficient of $S_1 s_2$  to vanish due to material symmetry requirements. Finally, using Eqs.~(\ref{eq:willis_asym_pm}, \ref{eq:s1s2_coeff}) along with the macro-scale Eq.~\eqref{eq:heq} one could determine the homogenized parameters as follows,
	\begin{equation}
		\begin{aligned}
			\eta_{33} &= -\dfrac{2}{\Delta},\\
			\mu_{55} &= \dfrac{4S_1^+Z_{53}^+Z_{53}^- + s_2(T_4^+ - T_4^-)(Z_{53}^+ - Z_{53}^-)}{2S_1^+\Delta},\\
			\mu_{44} &= \dfrac{4s_2T_4^+T_4^- + S_1^+(T_4^+ - T_4^-)(Z_{53}^+ - Z_{53}^-)}{2s_2\Delta},\\
			\mu_{45} &= -\dfrac{(T_4^+ + T_4^-)(Z_{53}^+ + Z_{53}^-)}{2\Delta},\\
			\kap_{34} &= \dfrac{T_4^+ + T_4^-}{\Delta},\\
			\kap_{35} &= -\dfrac{Z_{53}^+ + Z_{53}^-}{\Delta},
		\end{aligned}
	\end{equation}
	where $Z_{53}^\pm = -T_5^\pm/V_3^\pm$ and $\Delta = S_1^+(Z_{53}^- - Z_{53}^+) + s_2(T_4^+ - T_4^-)$. These results are valid for asymmetric and symmetric cells with fully elastic or viscoelastic layers. For symmetric cells it can be shown that,
	\begin{equation}
		\begin{aligned}
			\eta_{33} &= \dfrac{V_3^\pm}{P_3^\pm},&  \mu_{44} &= \dfrac{T_4^\pm}{\Gamma_4^\pm},&  \mu_{55} &= \dfrac{T_5^\pm}{\Gamma_5^\pm}, &(sym.\ str.),& \\
			\mu_{54} &= \mu_{45} = 0,&  \kap_{43} &= -\kap_{34} = 0,&  \kap_{53} &= -\kap_{35} = 0, &(sym.\ str.).&
		\end{aligned}
	\end{equation}
	The vanishing of the coupling parameters for symmetric unit cells is easily observed from $Z_{53}^- = -Z_{53}^+$, and $T_4^-= -T_4^+$. 
	\section{Illustrative examples: Overall constitutive Willis parameters from integration scheme and comparison with scattering}
\begin{figure}[!ht]
		\centering\includegraphics[height=120pt,center]{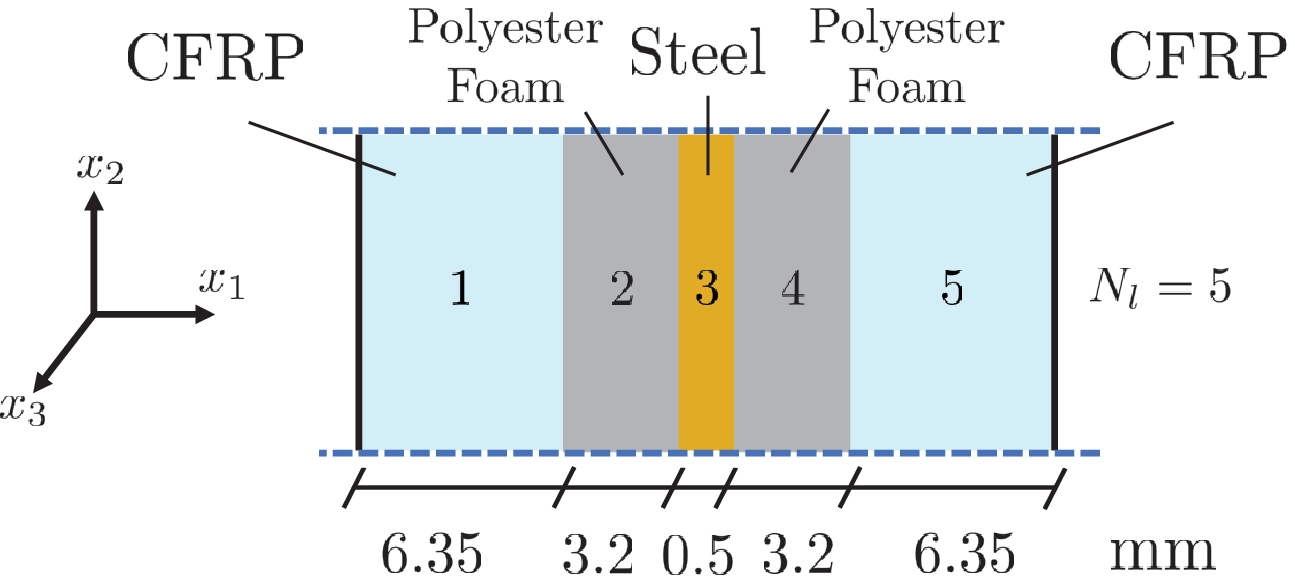}
		\caption{\label{fig:mat} The baseline 3-phase repeating unit cell for the layered media studied in this paper. Each unit cells consists of $N_l = 5$ layers. For scattering analysis two semi-infinite homogeneous polycarbonate layers are assumed on either side of the finite specimen with $N_c$ repeating unit cells. Propagation and scattering of oblique anti-plane shear (SH with respect to $x_3$ axis) incident waves are studied. Mechanical properties for each layer including, Young's modulus, density, shear and longitudinal waves speeds are the same as in \cite{Nemat-Nasser2015} and are summarized here: $(d^j) = (6.35, 3.20, 0.50, 3.20, 6.35)$, $(c’^j_T) = (1.04, 0.13, 2.88, 0.13, 1.04$, $(\rho^j) = (1.53, 0.36, 7.82, 0.36, 1.53)$. Note that for the polyester foam layers ($j = 2, 4$) 5\% loss is considered in shear wave speeds ($c_T''/c_T' = 0.05$, where $c_T = c_T' + \ii c_T''$).}
	\end{figure}

	Fig.~(\ref{fig:mat}) shows a baseline repeating unit cell of a sample 1D layered medium studied here for its scattering, band structure, and homogenized properties. In all the future graphs, the unit system [mm, $\mu$s, mg] for length, time, and mass are used which results in [km/s, GPa, g/cm$^{3}$, MHz, MRayl] units for wave speed (and inverse slowness), modulus, density (and inverse specific volume), frequency, and impedance, respectively. Frequency scale has been changed to kHz for clarity of graphical presentation. 
	
\begin{figure}[!ht]
		\begin{subfigure}[b]{\linewidth}
			\centering\includegraphics[height=205pt,center]{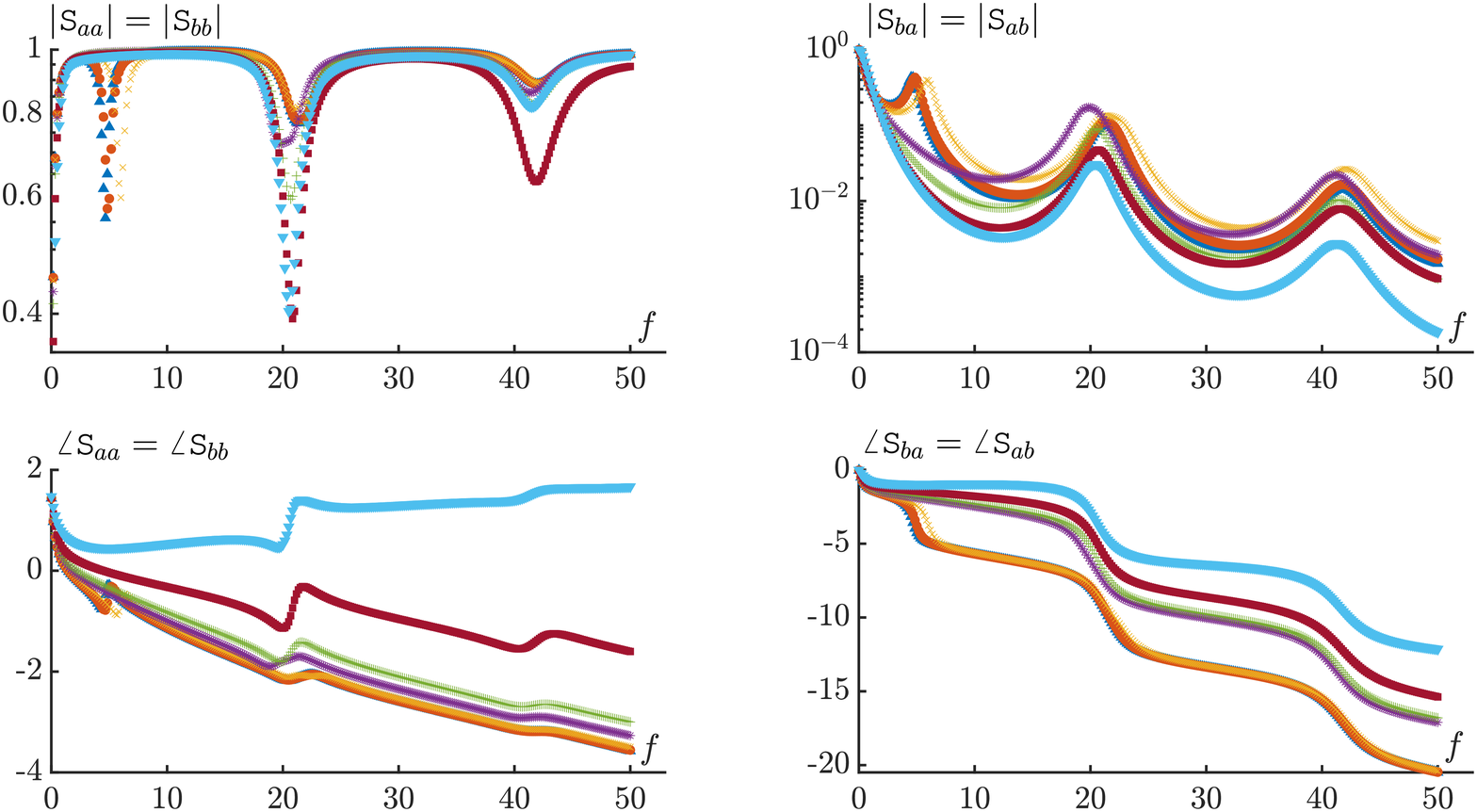}
			\caption{\label{fig:scatt_sh}}
		\end{subfigure}
		\begin{subfigure}[b]{\linewidth}
			\centering\includegraphics[height=140pt,center]{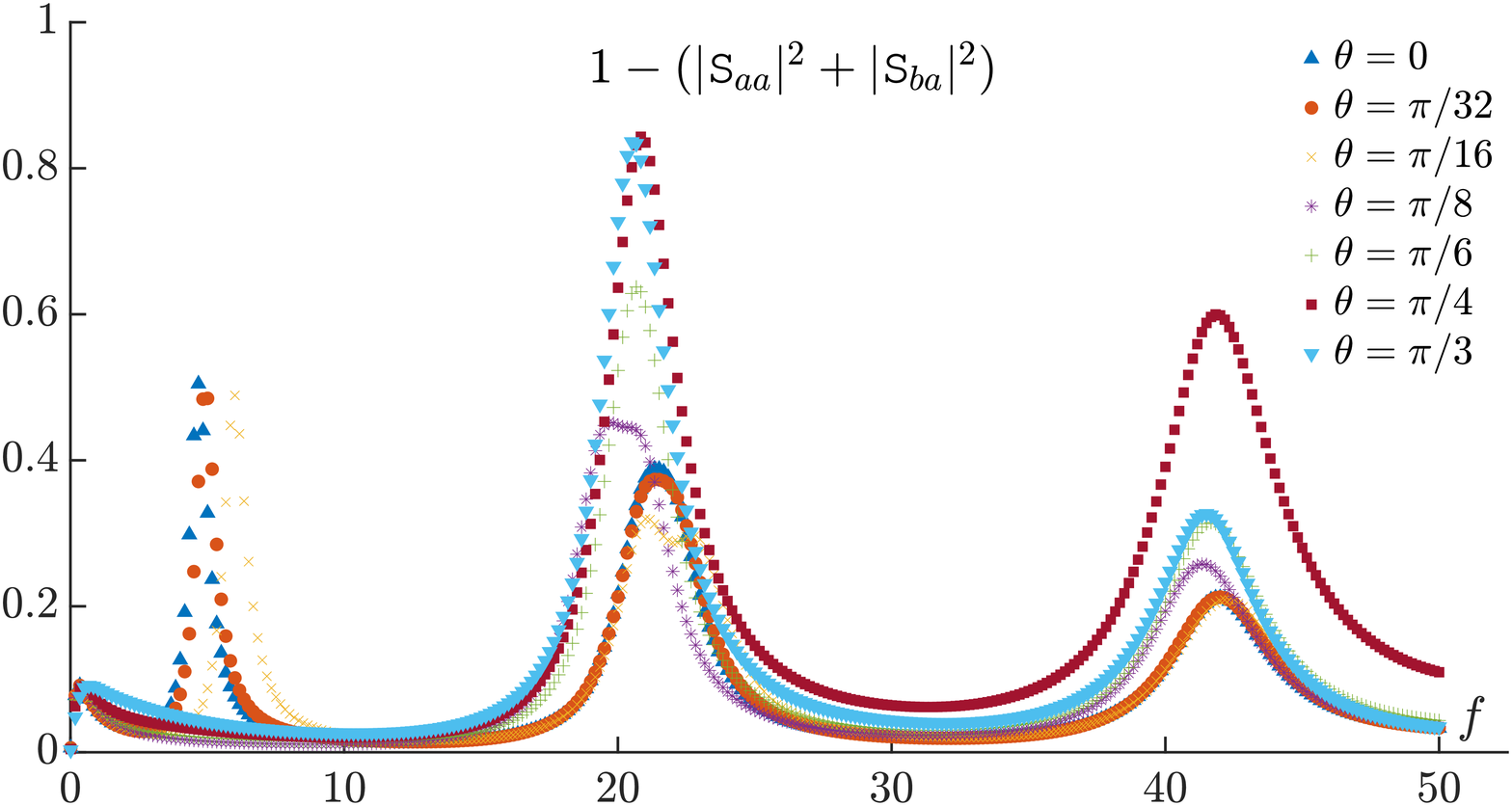}
			\caption{\label{fig:scatt_absp_sh}}
		\end{subfigure}
		\caption{\label{fig:scatt} Reflections and transmission magnitude and phase for the symmetric 5-layer RUC shown in Fig.~(\ref{fig:mat}) with one unit cell ($N_c = 1$). (\subref{fig:scatt_sh}) Top row shows the magnitude in log scale. It should be noted that at the low frequencies, there is almost full transmission and as the frequency increases, transmission reduces substantially. For higher angles of incident, transmission amplitude is lower. 
			Second row shows the phase angle of the reflection and transmission. (\subref{fig:scatt_absp_sh}) Bottom figure demonstrates the absorption for the 5-layer medium. Fully elastic case should yield zero value for absorption due to energy conservation. For the lossy case analyzed here ($c_T''/c_T' = 0.05$ for polyester foam layer) absorption peaks are observed at the resonance frequencies.}
	\end{figure}

	\paragraph{Scattering} Scattering coefficients magnitudes and unwrapped phases for the 3-phase, 5-layer, RUC are shown as functions of incident angle and frequency in Fig.~(\ref{fig:scatt}). It should be noted that, since each constituent material is reciprocal, the whole medium is reciprocal as well, subsequently $\SM_{ba}=\SM_{ab}$ \cite{achenbach_2004}. Moreover, due to the parity symmetry of this particular design, $\SM_{aa}=\SM_{bb}$ (see \ref{app:scatt}). 
	For a similar analysis on a 2-phase 3-layered medium, the reader is referred to \cite{Amirkhizi2018}). The graphs are drawn for scattering off of a single unit cell $N_c = 1$. The main features occur at close but not exactly the same frequencies for all angles of incidence and certain low frequency features are only observable at very low angles of incidence (near normal). 
	
\begin{figure}[!ht]
		\begin{subfigure}[b]{0.5\linewidth}
			\centering\includegraphics[height=115pt,center]{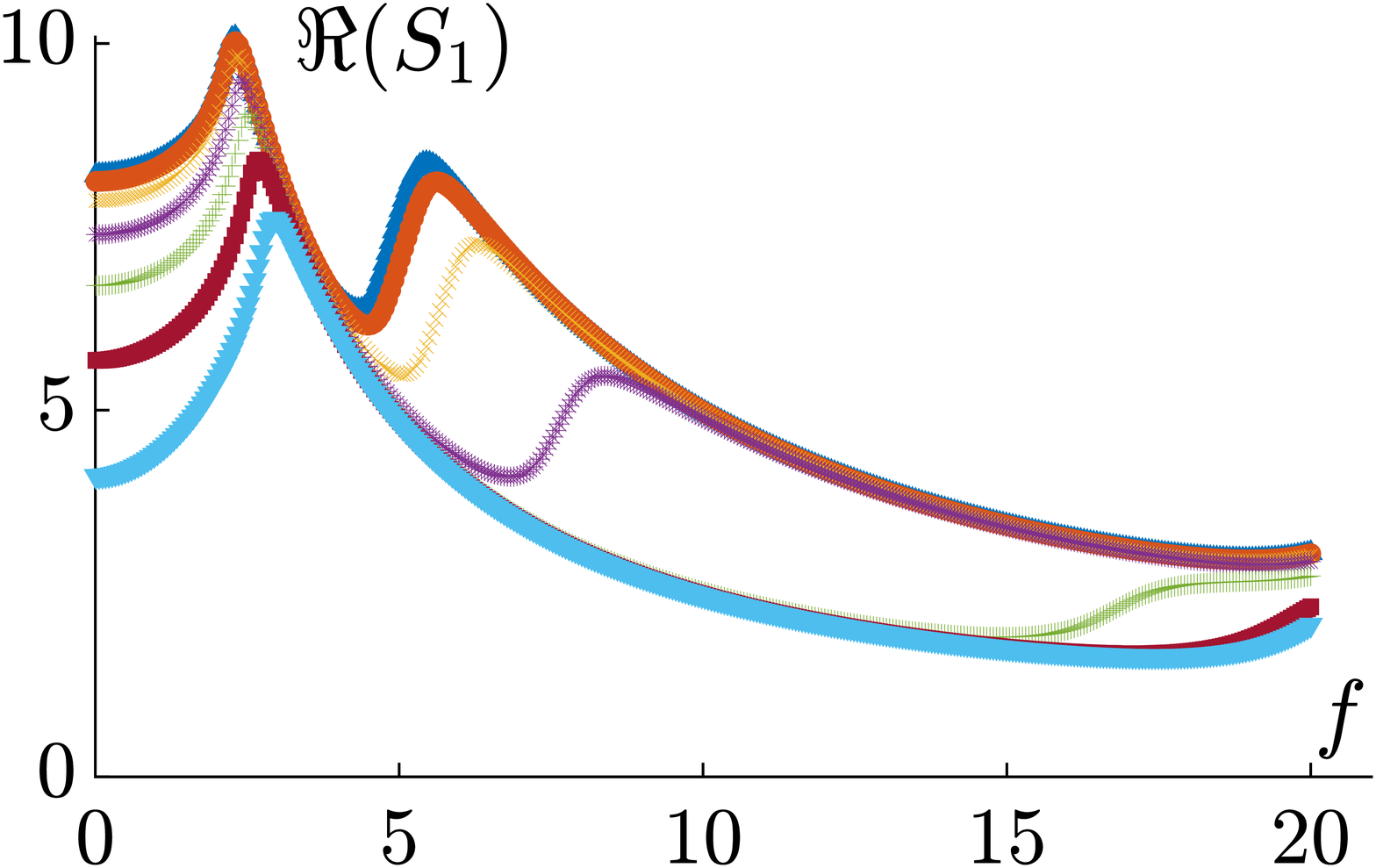}
			\caption{\label{fig:bs_re_s1_sh}}
		\end{subfigure}%
		\begin{subfigure}[b]{0.5\linewidth}
			\centering\includegraphics[height=115pt,center]{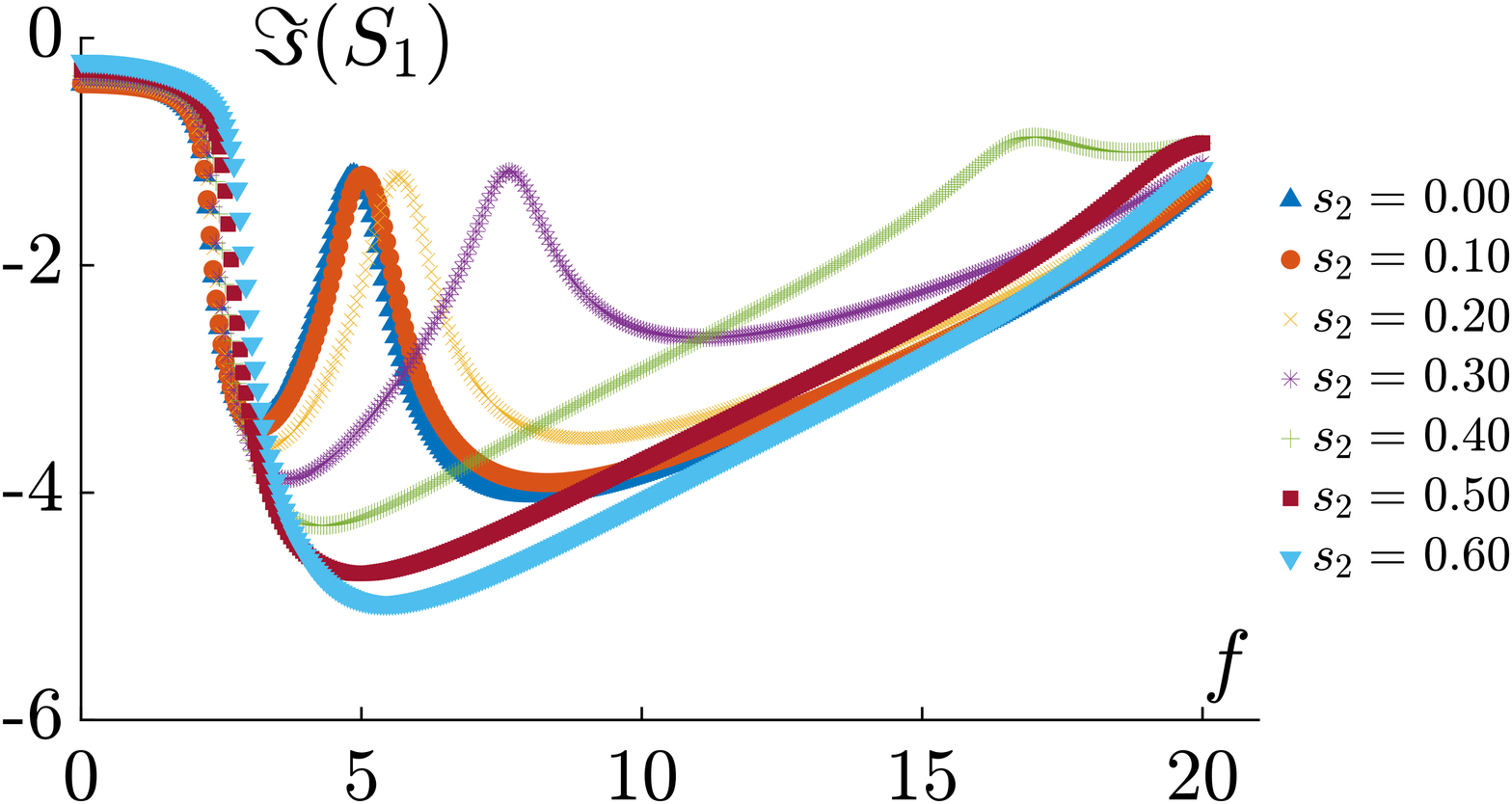}
			\caption{\label{fig:bs_im_s1_sh}}
		\end{subfigure}
		\begin{subfigure}[b]{0.5\linewidth}
			\centering\includegraphics[height=115pt,center]{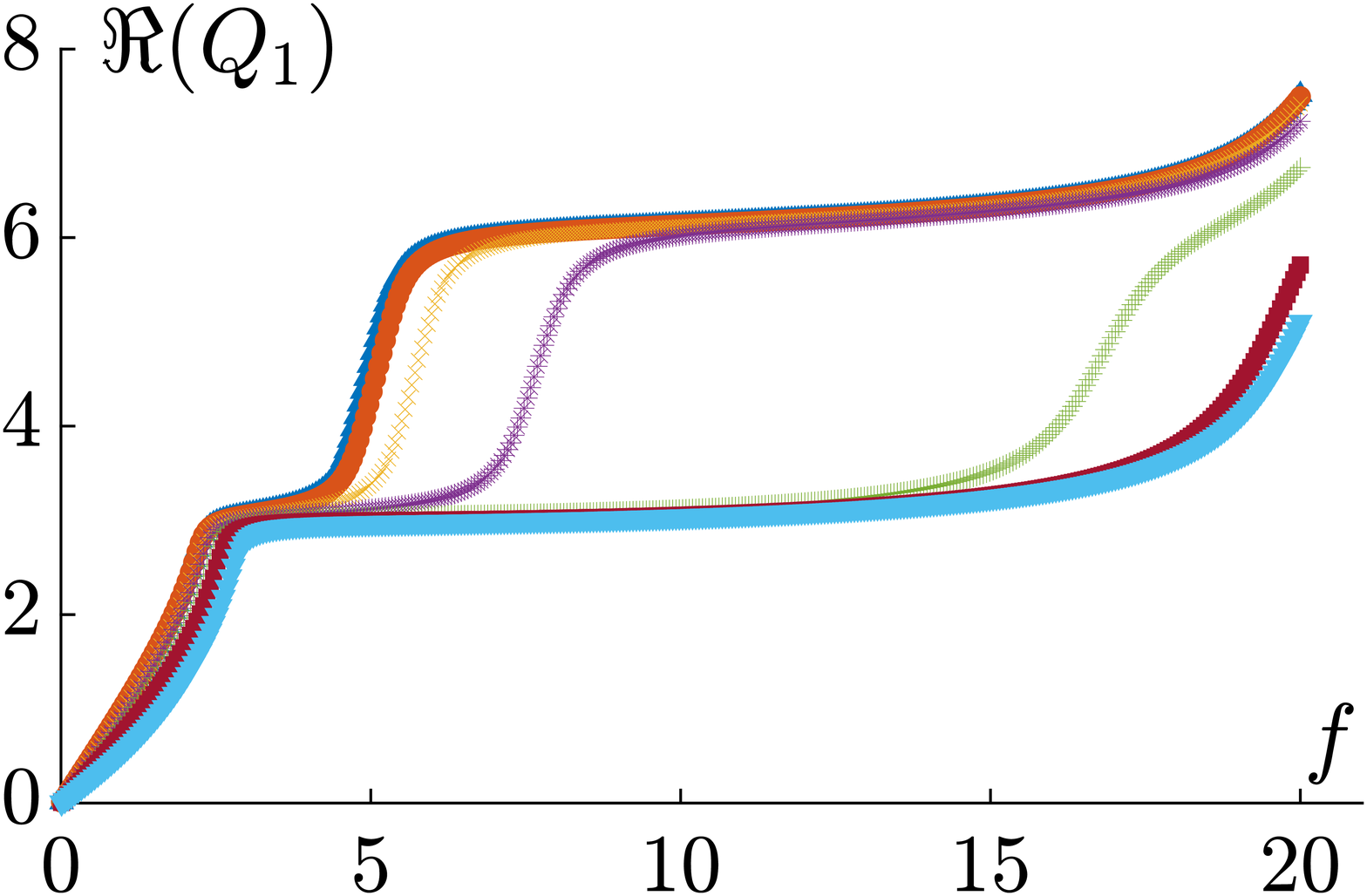}
			\caption{\label{fig:bs_re_q1_sh}}
		\end{subfigure}%
		\begin{subfigure}[b]{0.5\linewidth}
			\centering\includegraphics[height=115pt,center]{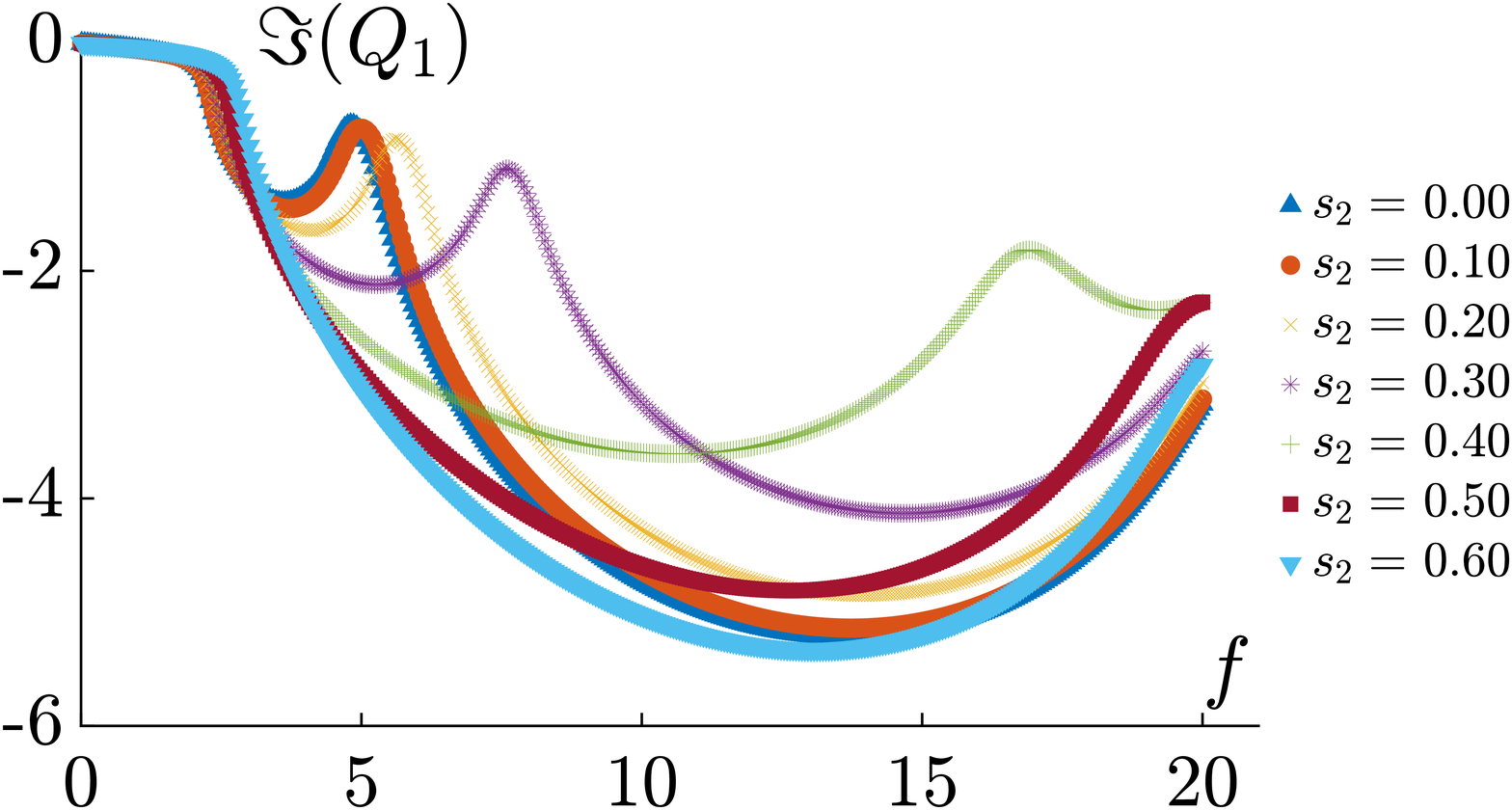}
			\caption{\label{fig:bs_im_q1_sh}}
		\end{subfigure}
		\caption{\label{fig:phs_adv} (\subref{fig:bs_re_s1_sh}) and (\subref{fig:bs_re_q1_sh}) show the real part of slowness and phase advance, respectively, for an infinitely periodic array with the 5-layer symmetric RUC shown in Fig.~(\ref{fig:mat}). (\subref{fig:bs_im_s1_sh}) and (\subref{fig:bs_im_q1_sh}) show their imaginary counterparts. Wide stop bands at relatively low frequencies (starting near 2.5kHZ) are observable for all $s_2$ values, and as $s_2$ increases stop bands become wider. The real part of the phase advance $Q_1$ is initially calculated to be between $-\pi$ and $\pi$, and by adding a suitable integer multiple of $2\pi$, it can be rendered a continuous function of frequency in such layered structures.}
	\end{figure}

	\paragraph{Band structure}
	The solution Eq.~\eqref{eq:tm-Q} of the eigenvalue problem Eq.~\eqref{eq:tm-eigen} with the transfer matrix of a unit cell for each frequency value yields the phase advance across a cell in an infinitely periodic array. Fig.~(\ref{fig:phs_adv}) shows the components of the slowness, $S_1 = K_1/ \omega$, and normalized wave vector, $Q_1 = K_1 d$ normal to the layers ($x_1$) as functions of frequency and $s_2 = k_2/\omega$, the component of the slowness parallel to the layers. $s_2 = 0$ represents normally incident waves. Symmetric and reciprocal unit cells lead to symmetric solutions for right- and left-traveling waves, i.e., $Q_1^- = -Q_1^+ = -Q_1$. The branch selection for Eq.~\eqref{eq:tm-Q} is done in order to maintain the continuity wave vector as a function of frequency. 
	
\begin{figure}[!ht]
		\begin{subfigure}[b]{0.5\linewidth}
			\centering\includegraphics[height=95pt,center]{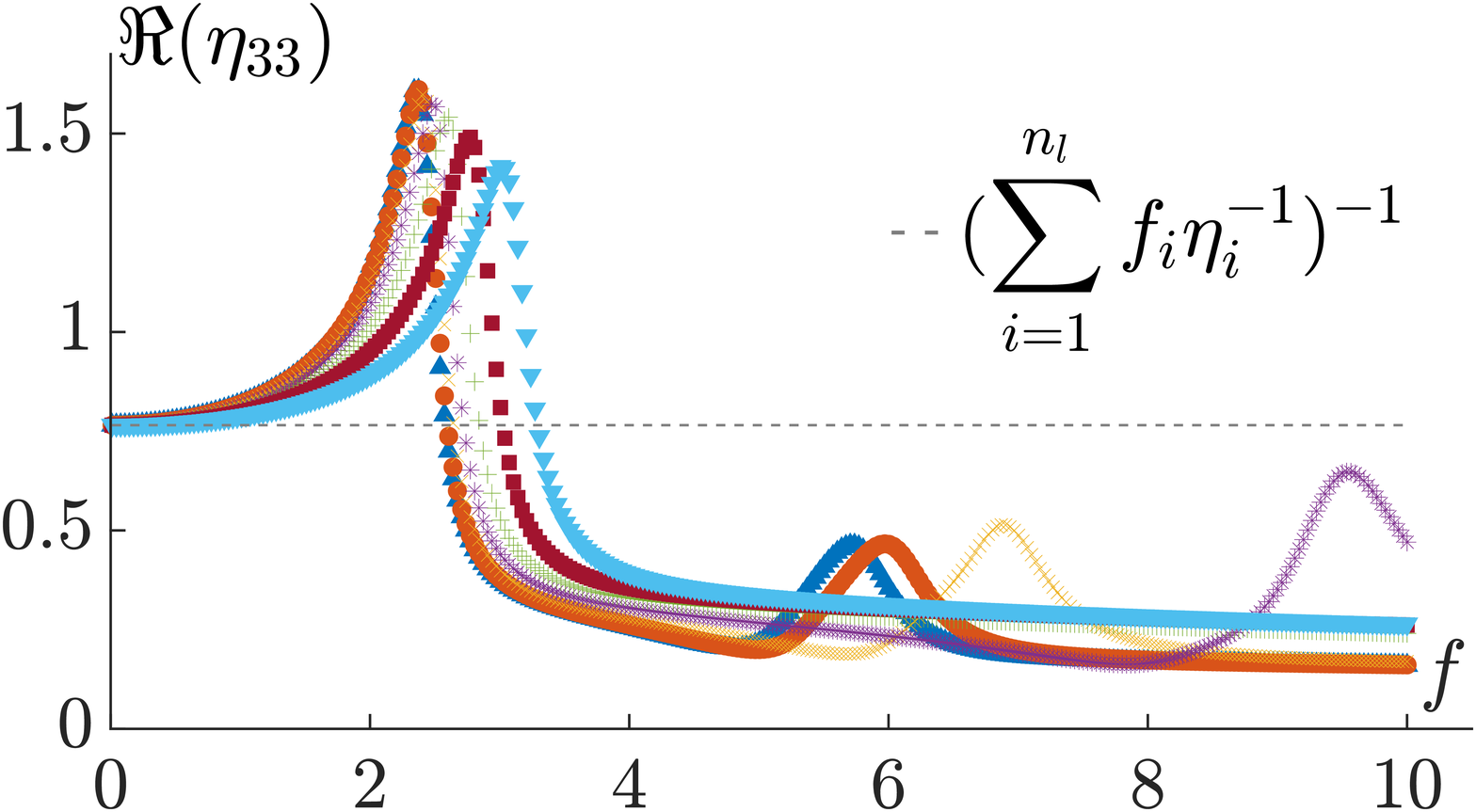}
			\caption{\label{fig:in_re_eta33_sh_asym_2}}
		\end{subfigure}%
		\begin{subfigure}[b]{0.5\linewidth}
			\centering\includegraphics[height=95pt,center]{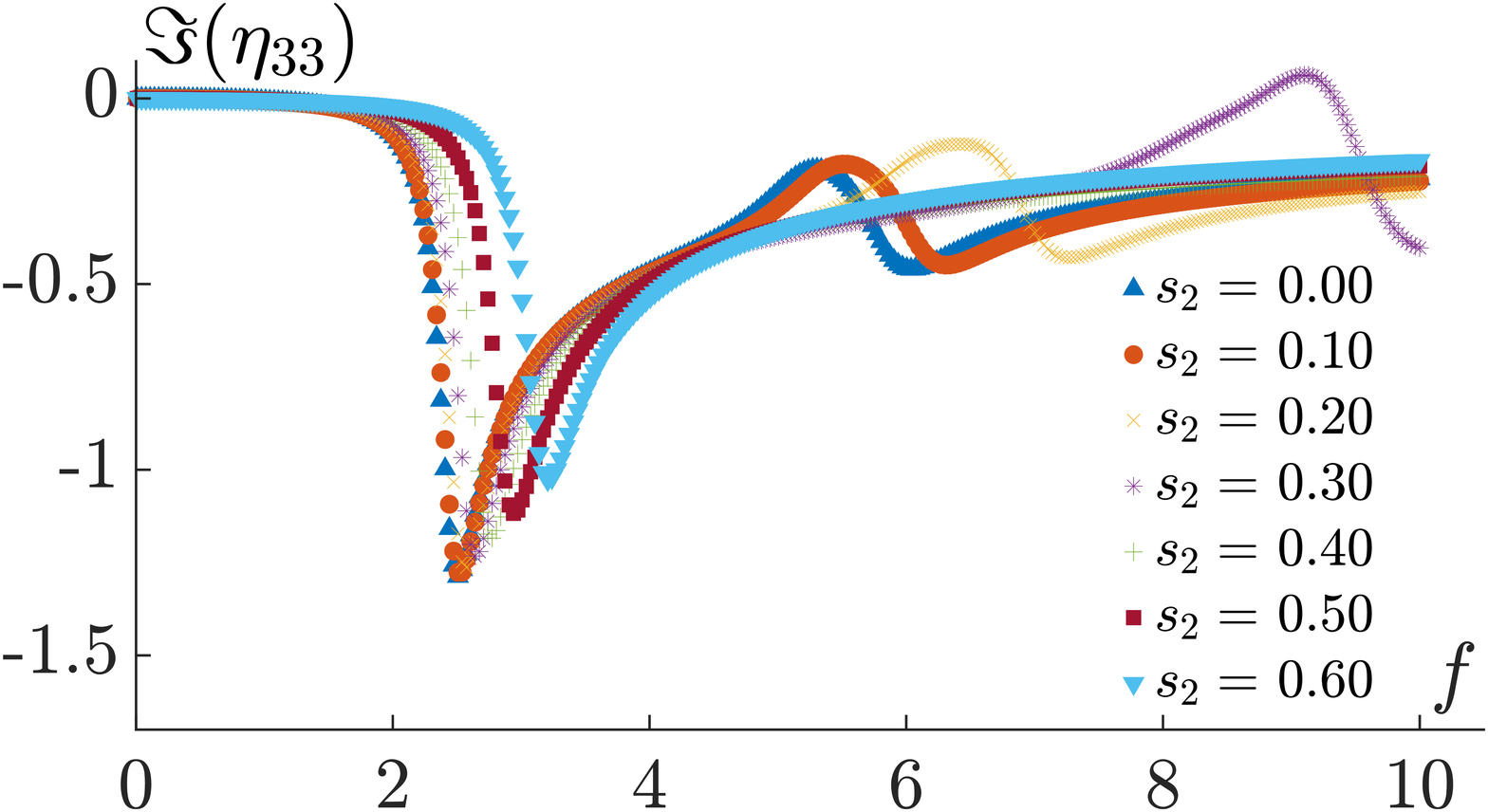}
			\caption{\label{fig:in_im_eta33_sh_asym_2}}
		\end{subfigure}
		\begin{subfigure}[b]{0.5\linewidth}
			\centering\includegraphics[height=95pt,center]{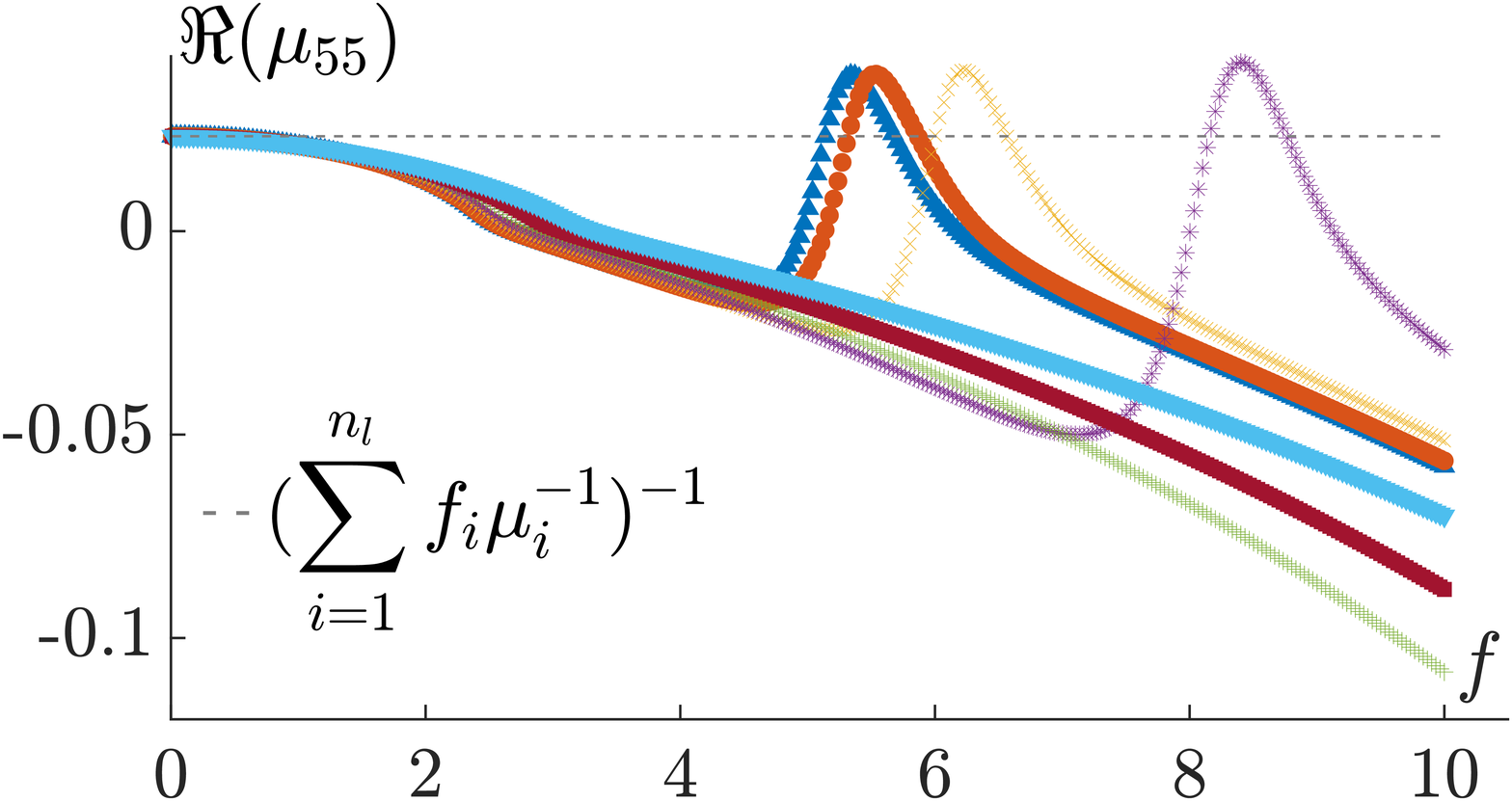}
			\caption{\label{fig:in_re_mu55_sh_asym_2}}
		\end{subfigure}%
		\begin{subfigure}[b]{0.5\linewidth}
			\centering\includegraphics[height=95pt,center]{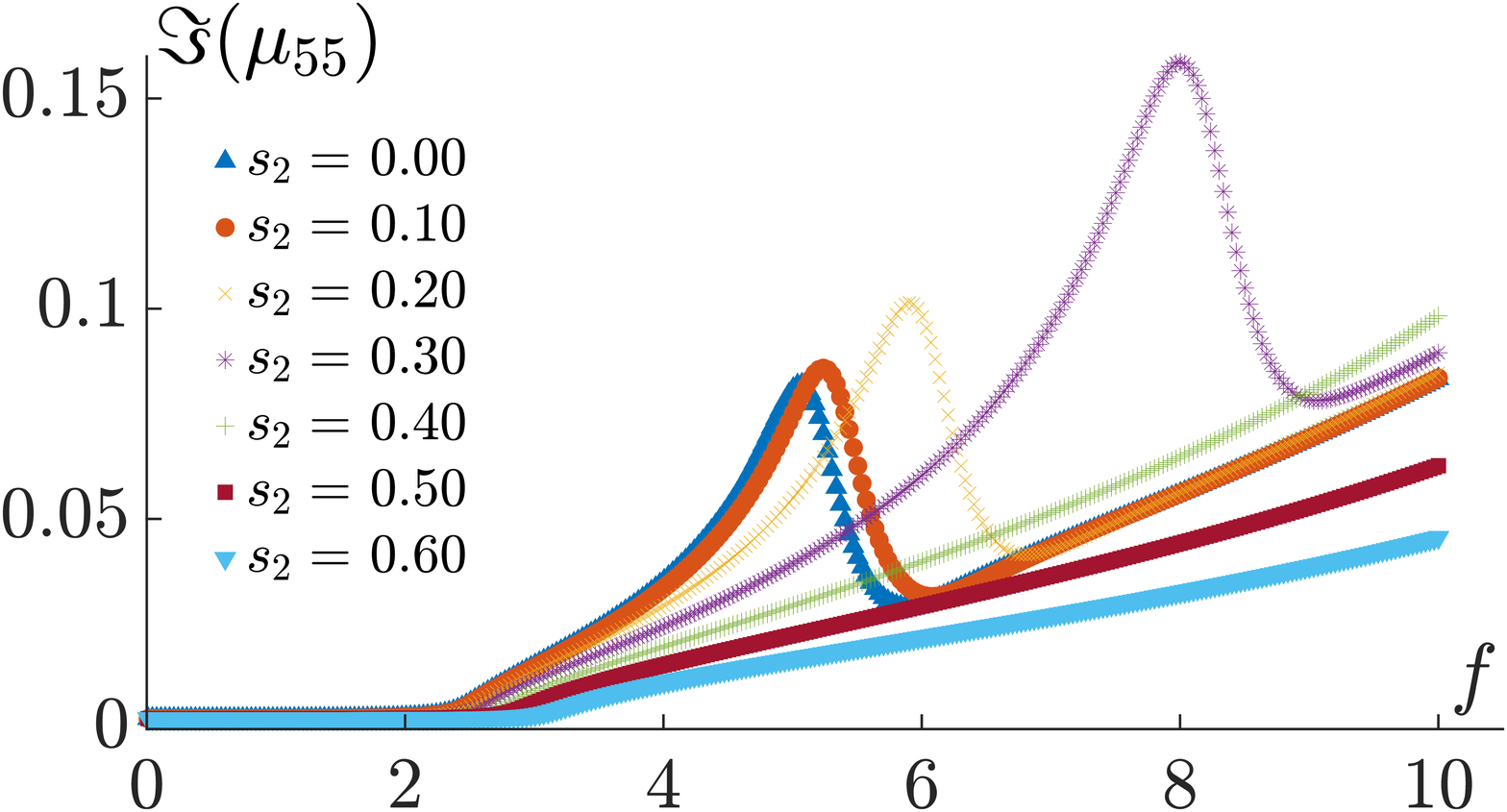}
			\caption{\label{fig:in_im_mu55_sh_asym_2}}
		\end{subfigure}
		\begin{subfigure}[b]{0.5\linewidth}
			\centering\includegraphics[height=95pt,center]{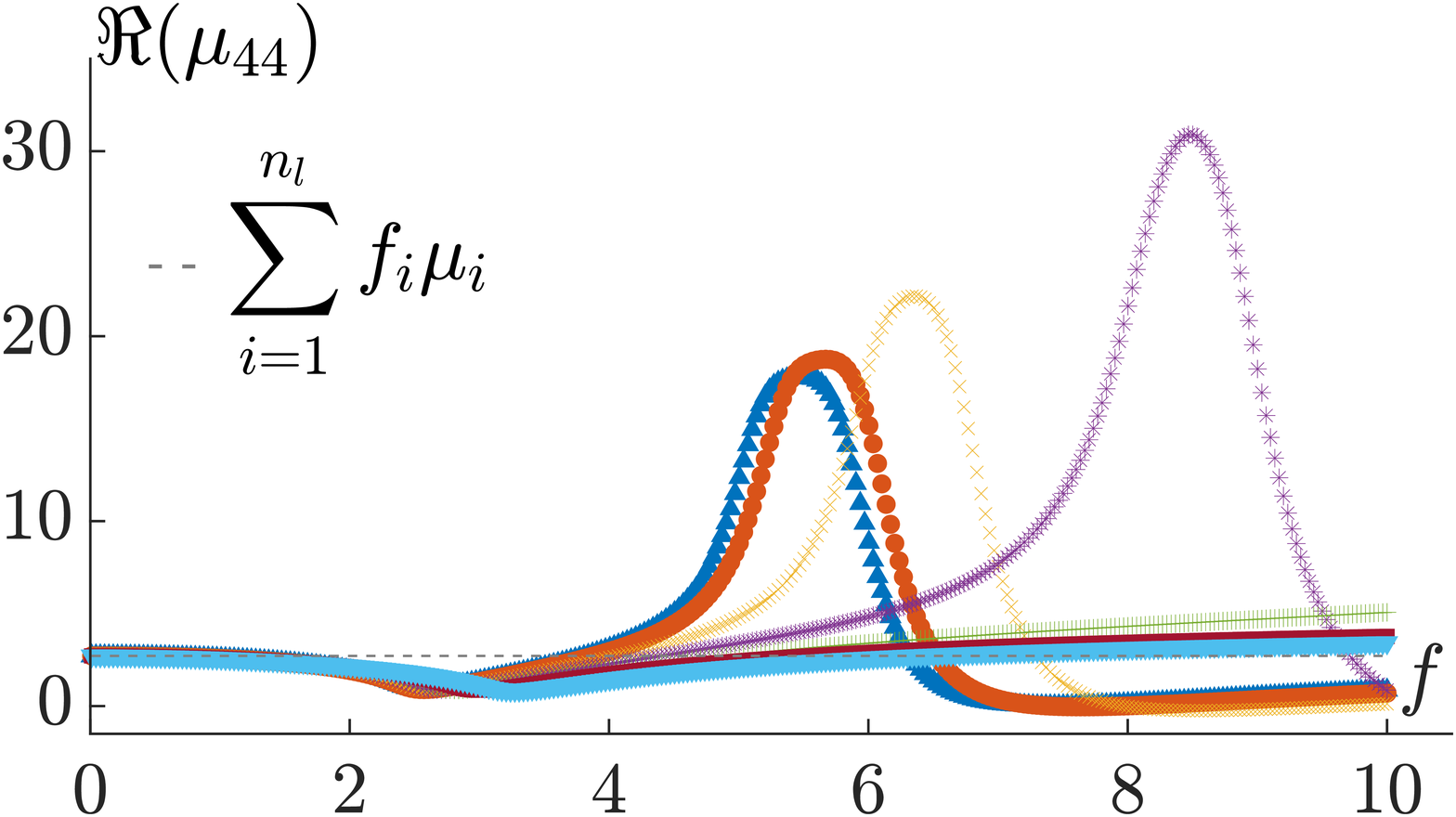}
			\caption{\label{fig:in_re_mu44_sh_asym_2}}
		\end{subfigure}%
		\begin{subfigure}[b]{0.5\linewidth}
			\centering\includegraphics[height=95pt,center]{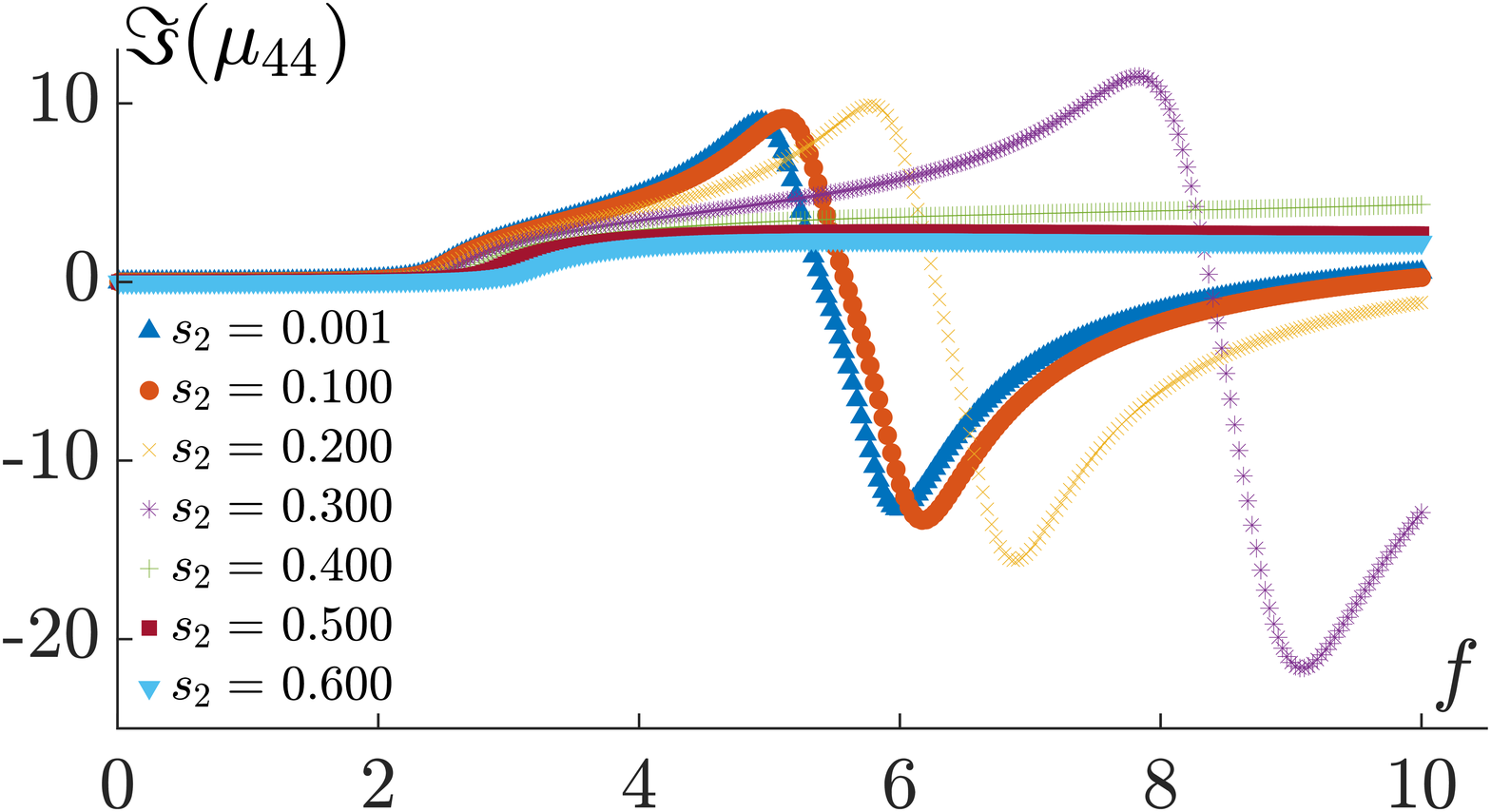}
			\caption{\label{fig:in_im_mu44_sh_asym_2}}
		\end{subfigure}
		\caption{\label{fig:asym_dispersive_diag} Homogenized diagonal components of the Willis constitutive tensor for an asymmetric RUC are shown here. (\subref{fig:in_re_eta33_sh_asym_2}), (\subref{fig:in_re_mu55_sh_asym_2}), and (\subref{fig:in_re_mu44_sh_asym_2}) show the real components of $\eta_{33}$, $\mu_{55}$, and $\mu_{44}$, and (\subref{fig:in_im_eta33_sh_asym_2}), (\subref{fig:in_im_mu55_sh_asym_2}), and (\subref{fig:in_im_mu44_sh_asym_2}) show their imaginary counterparts, respectively. The dependence on wave vector is observable for all these quantities. The real components of the diagonal terms are converging to their relevant Voigt and Reuss averages at long-wavelength limit for all wave directions.}
	\end{figure}

\begin{figure}[!ht]
		\begin{subfigure}[b]{0.5\linewidth}
			\centering\includegraphics[height=95pt,center]{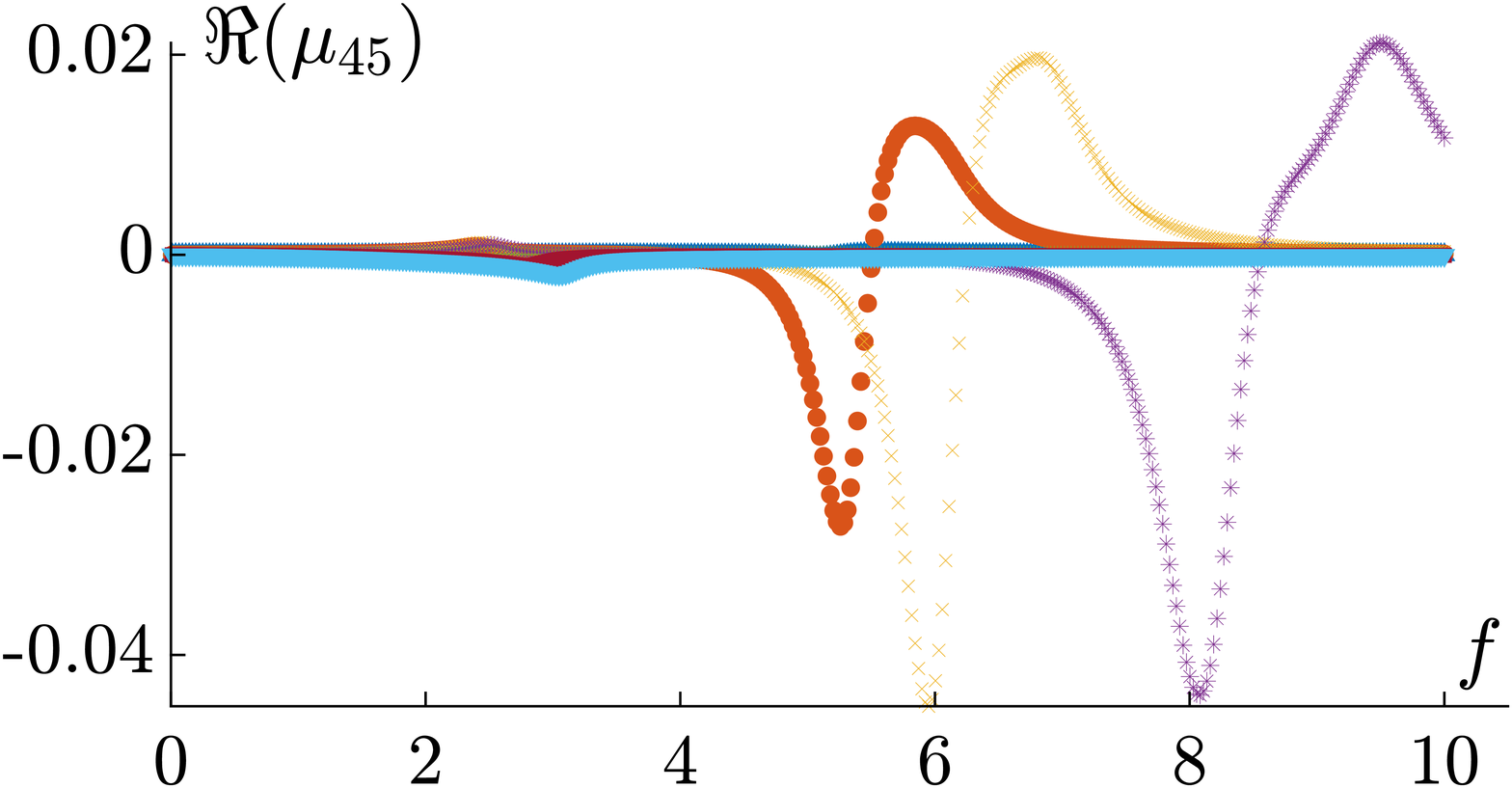}
			\caption{\label{fig:in_re_mu45_sh_asym_2}}
		\end{subfigure}%
		\begin{subfigure}[b]{0.5\linewidth}
			\centering\includegraphics[height=95pt,center]{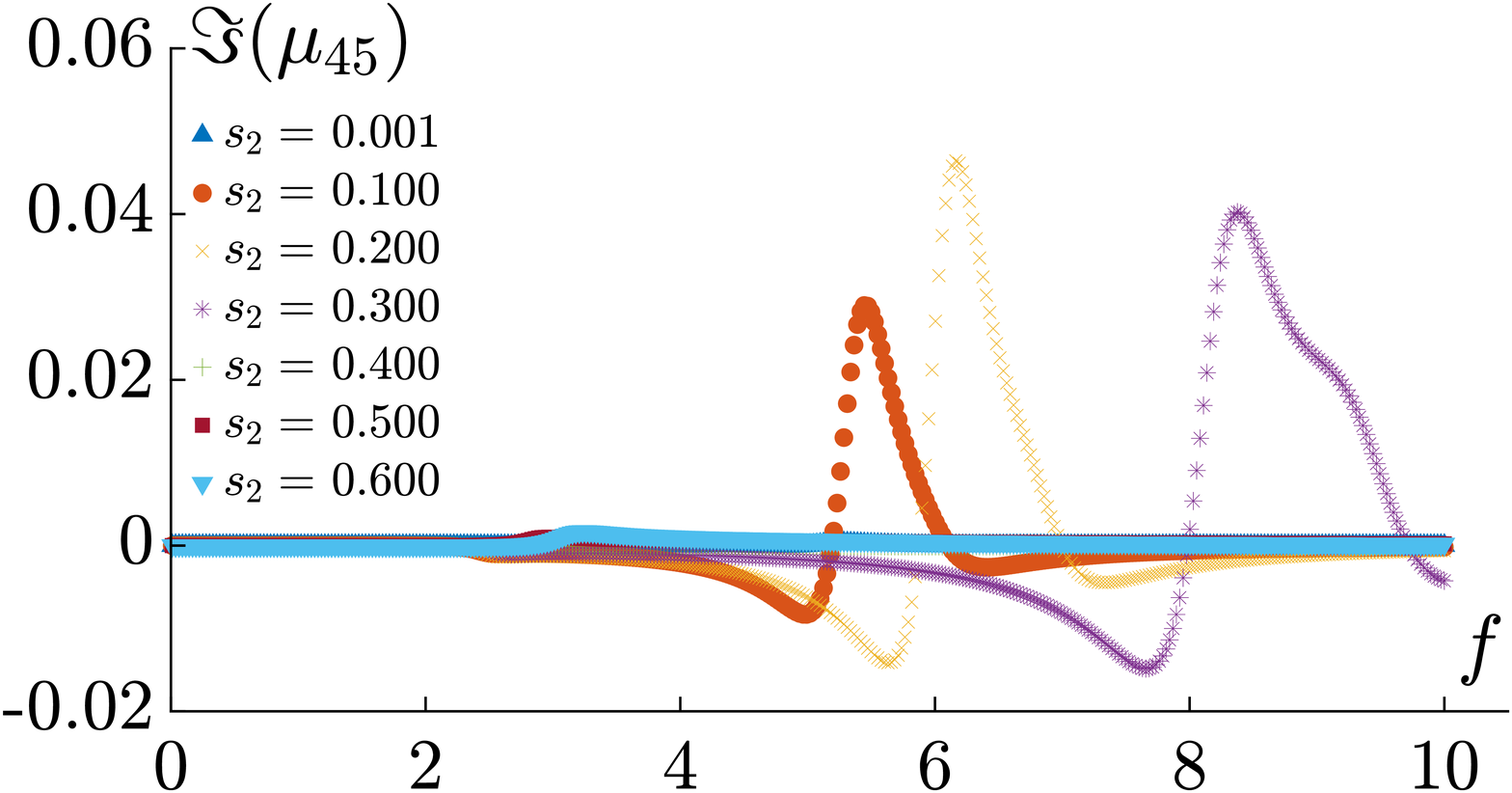}
			\caption{\label{fig:in_im_mu45_sh_asym_2}}
		\end{subfigure}
		\begin{subfigure}[b]{0.5\linewidth}
			\centering\includegraphics[height=95pt,center]{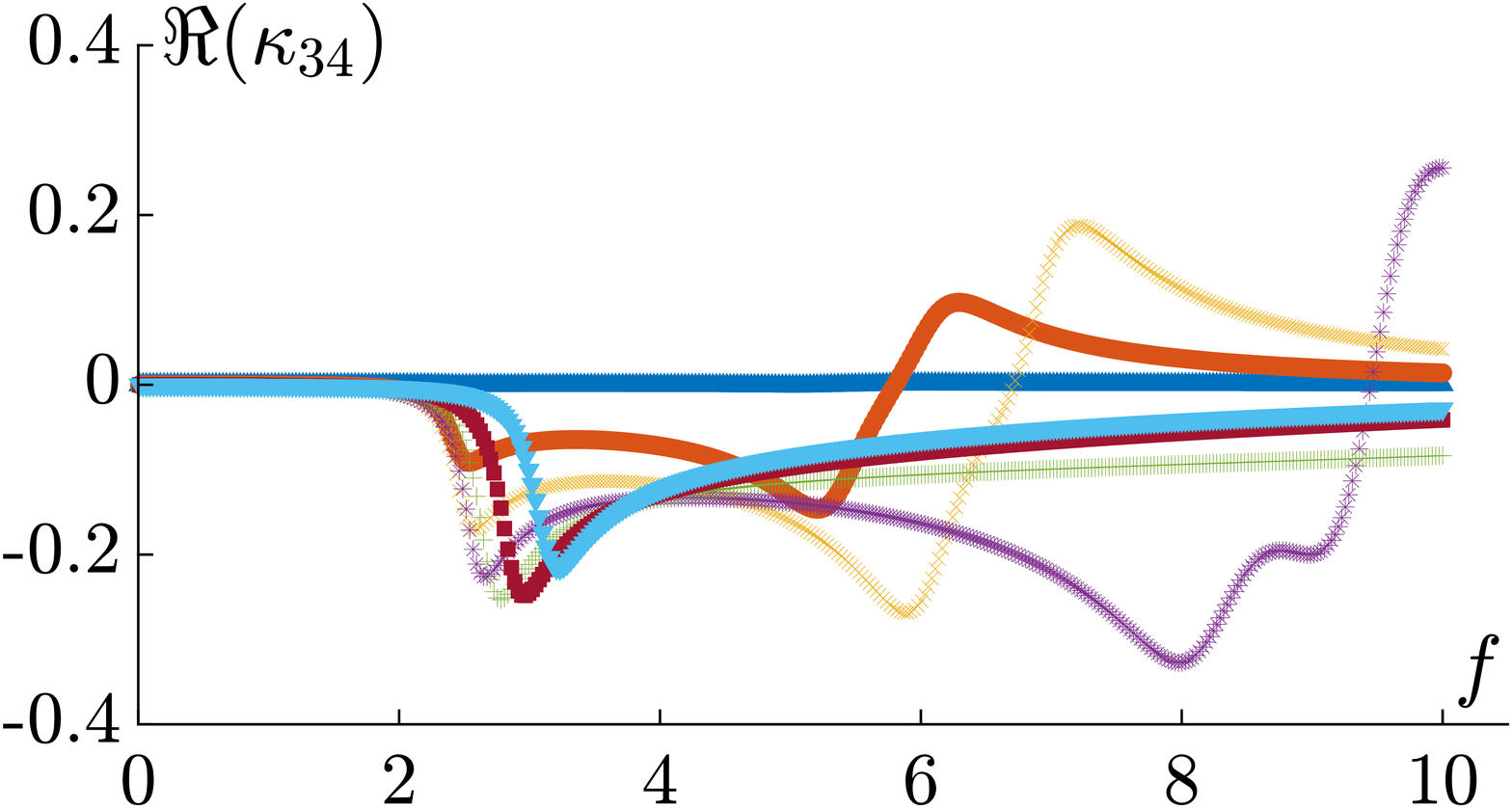}
			\caption{\label{fig:in_re_kap34_sh_asym_2}}
		\end{subfigure}%
		\begin{subfigure}[b]{0.5\linewidth}
			\centering\includegraphics[height=95pt,center]{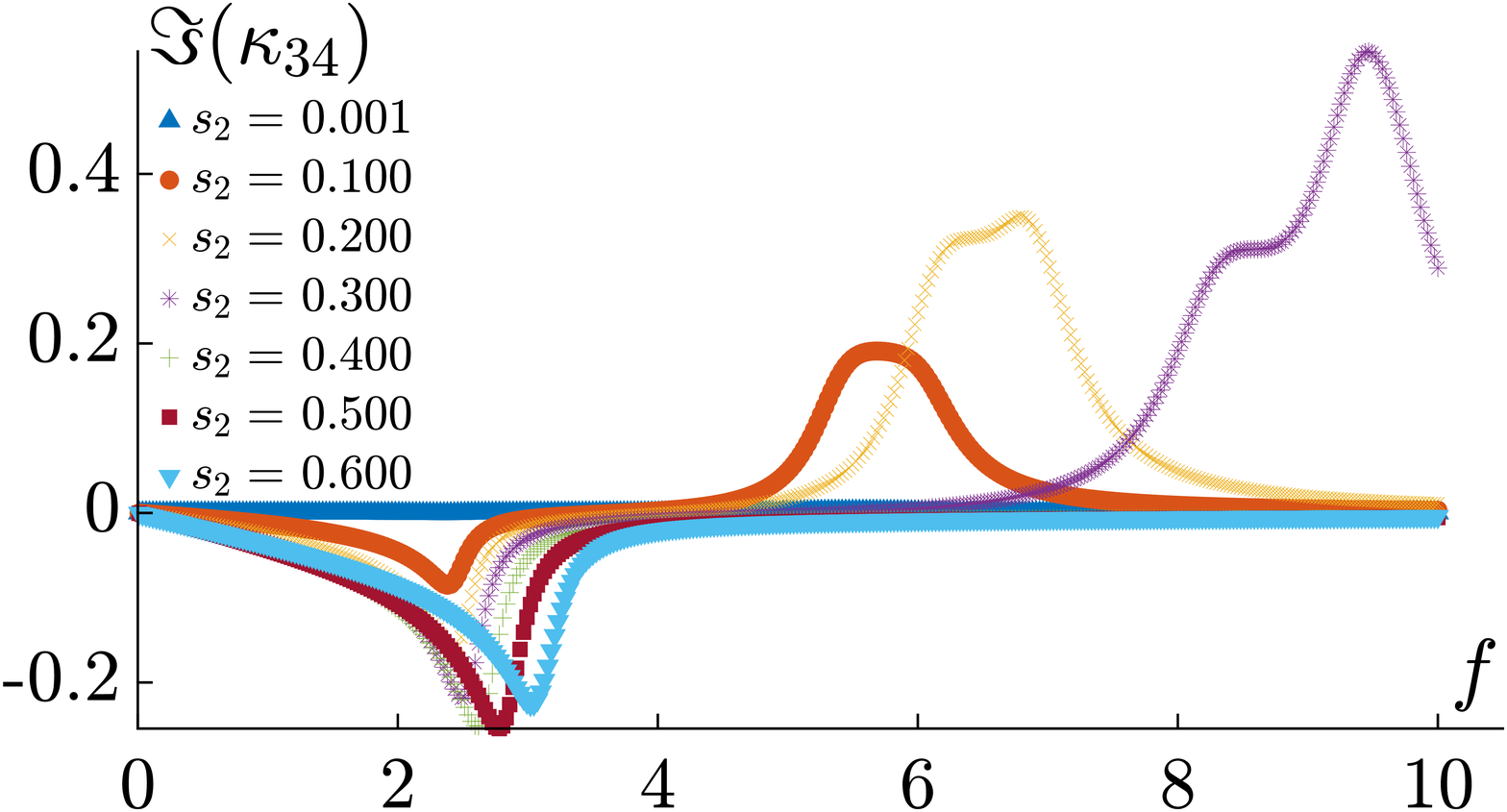}
			\caption{\label{fig:in_im_kap34_sh_asym_2}}
		\end{subfigure}
		\begin{subfigure}[b]{0.5\linewidth}
			\centering\includegraphics[height=95pt,center]{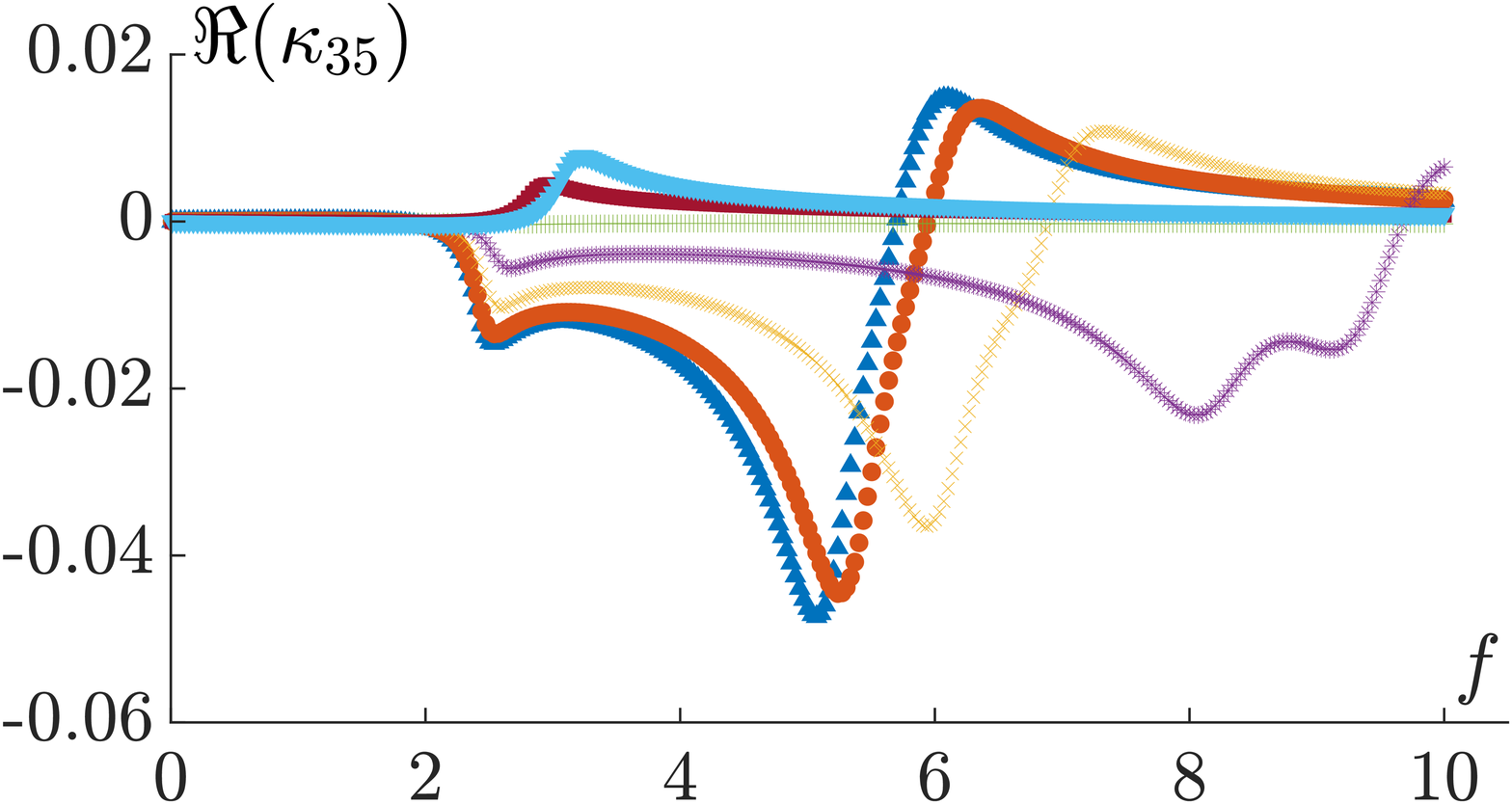}
			\caption{\label{fig:in_re_kap35_sh_asym_2}}
		\end{subfigure}%
		\begin{subfigure}[b]{0.5\linewidth}
			\centering\includegraphics[height=95pt,center]{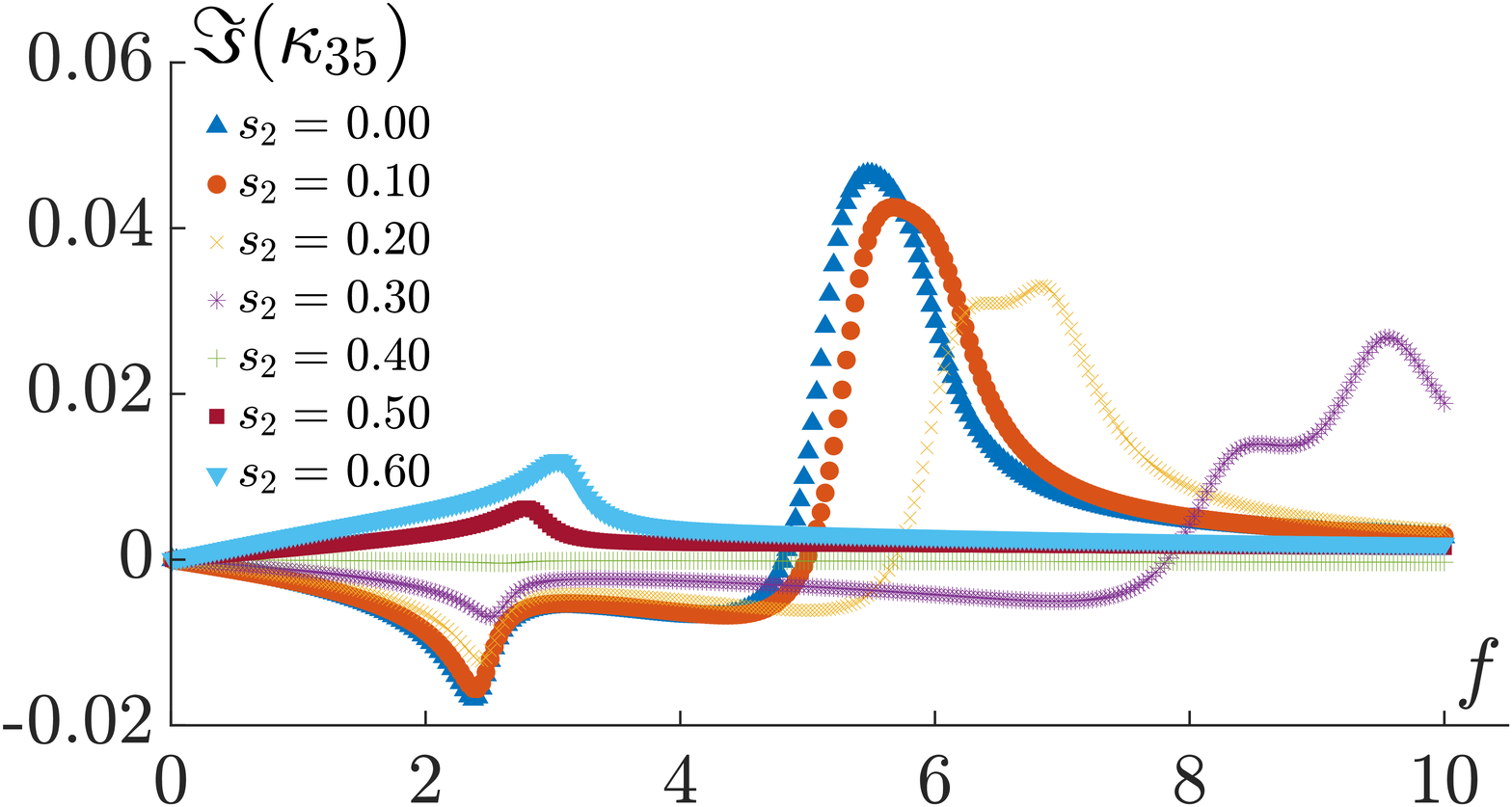}
			\caption{\label{fig:in_im_kap35_sh_asym_2}}
		\end{subfigure}
		\caption{\label{fig:asym_dispersive_off_diagonal} Homogenized off-diagonal components of the Willis constitutive tensor for an asymmetric cell. The real parts of the off-diagonal terms of $\mu_{45}$, $\kap_{34}$, and $\kap_{35}$ are shown in (\subref{fig:in_re_mu45_sh_asym_2}), (\subref{fig:in_re_kap34_sh_asym_2}), (\subref{fig:in_re_kap35_sh_asym_2}), while their imaginary counterparts are shown in (\subref{fig:in_im_mu45_sh_asym_2}), (\subref{fig:in_im_kap34_sh_asym_2}), and (\subref{fig:in_im_kap35_sh_asym_2}), respectively. Note that for the off-diagonal terms both real and imaginary parts approach zero as frequency decreases. Also note that for this particular example $\mu_{45}$, $\kap_{35}$ are an order of magnitude smaller than $\kap_{34}$. All the off-diagonal terms converge to zero when the RUC becomes fully symmetric.}
	\end{figure}

\begin{figure}[!ht]
		\begin{subfigure}[b]{0.5\linewidth}
			\centering\includegraphics[height=95pt,center]{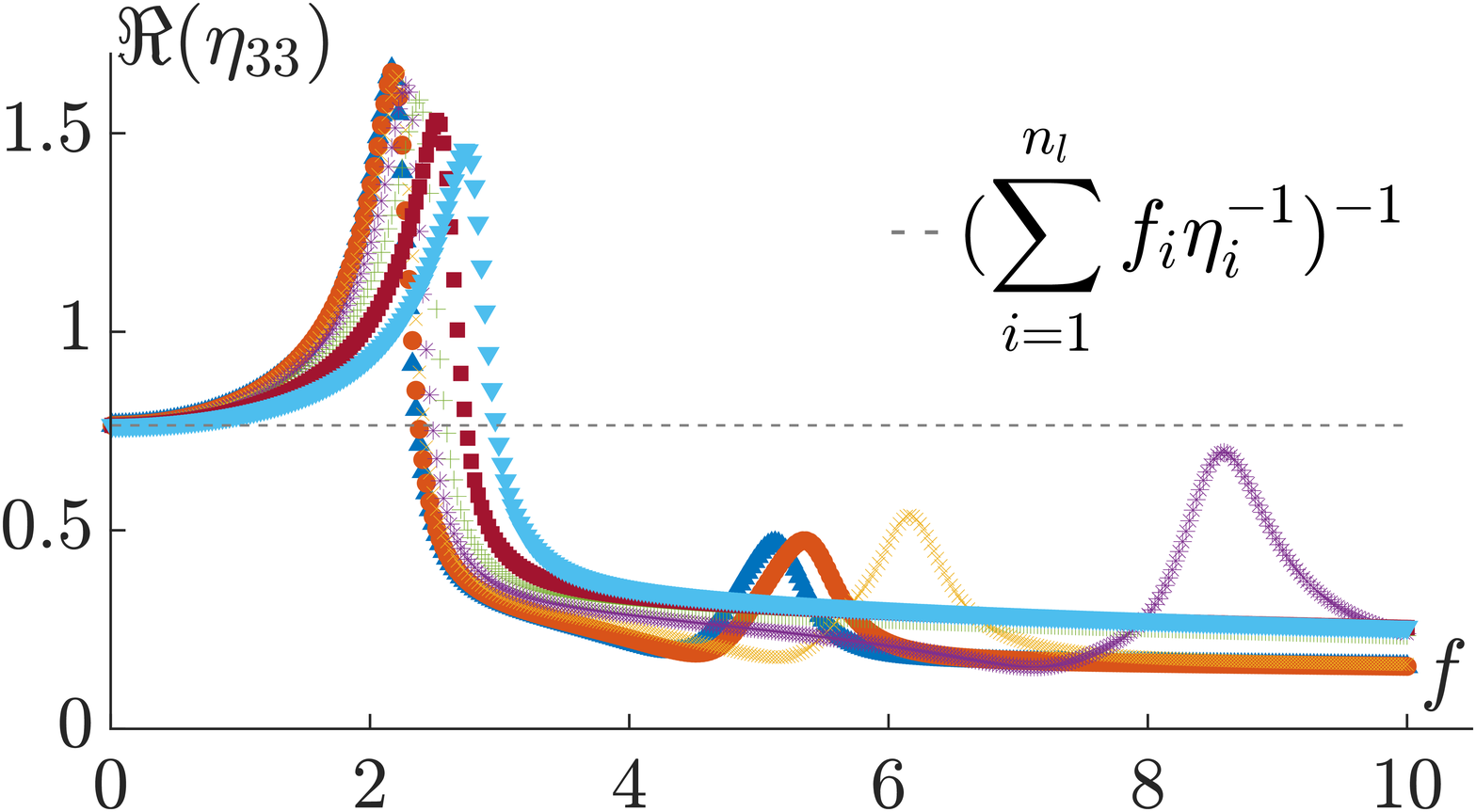}
			\caption{\label{fig:in_re_eta33_sh_sym_2}}
		\end{subfigure}%
		\begin{subfigure}[b]{0.5\linewidth}
			\centering\includegraphics[height=95pt,center]{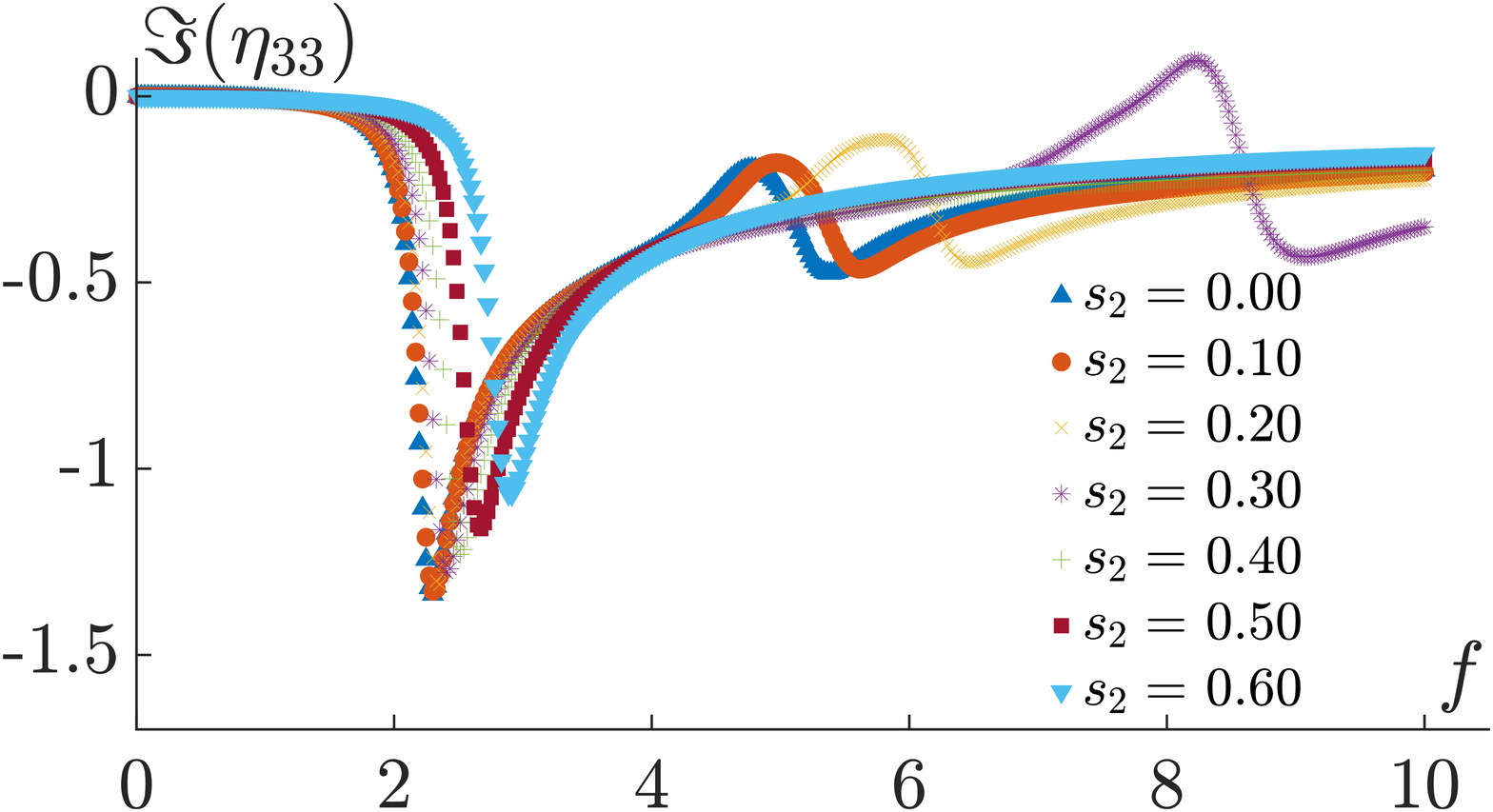}
			\caption{\label{fig:in_im_eta33_sh_sym_2}}
		\end{subfigure}
		\begin{subfigure}[b]{0.5\linewidth}
			\centering\includegraphics[height=95pt,center]{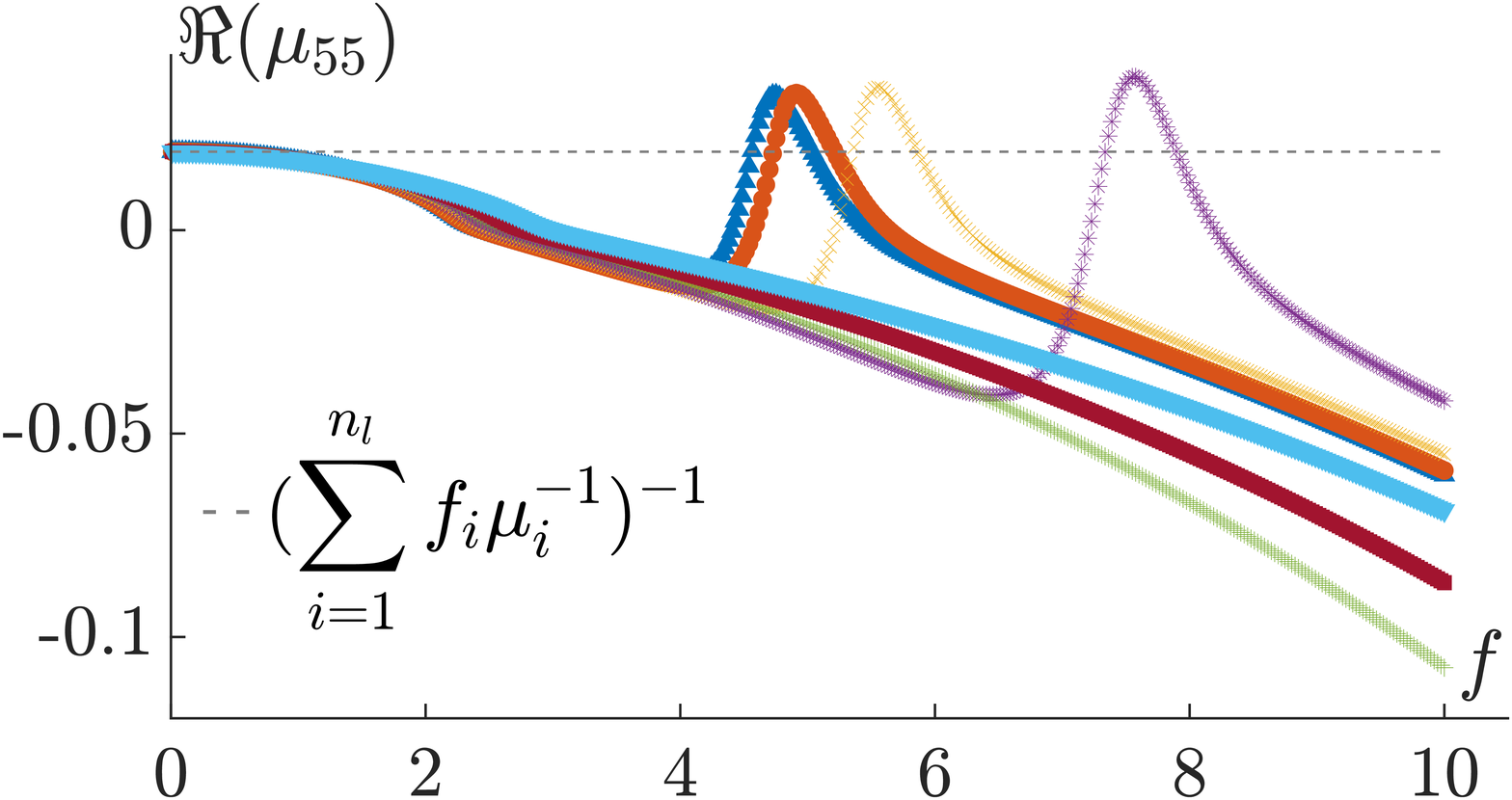}
			\caption{\label{fig:in_re_mu55_sh_sym_2}}
		\end{subfigure}%
		\begin{subfigure}[b]{0.5\linewidth}
			\centering\includegraphics[height=95pt,center]{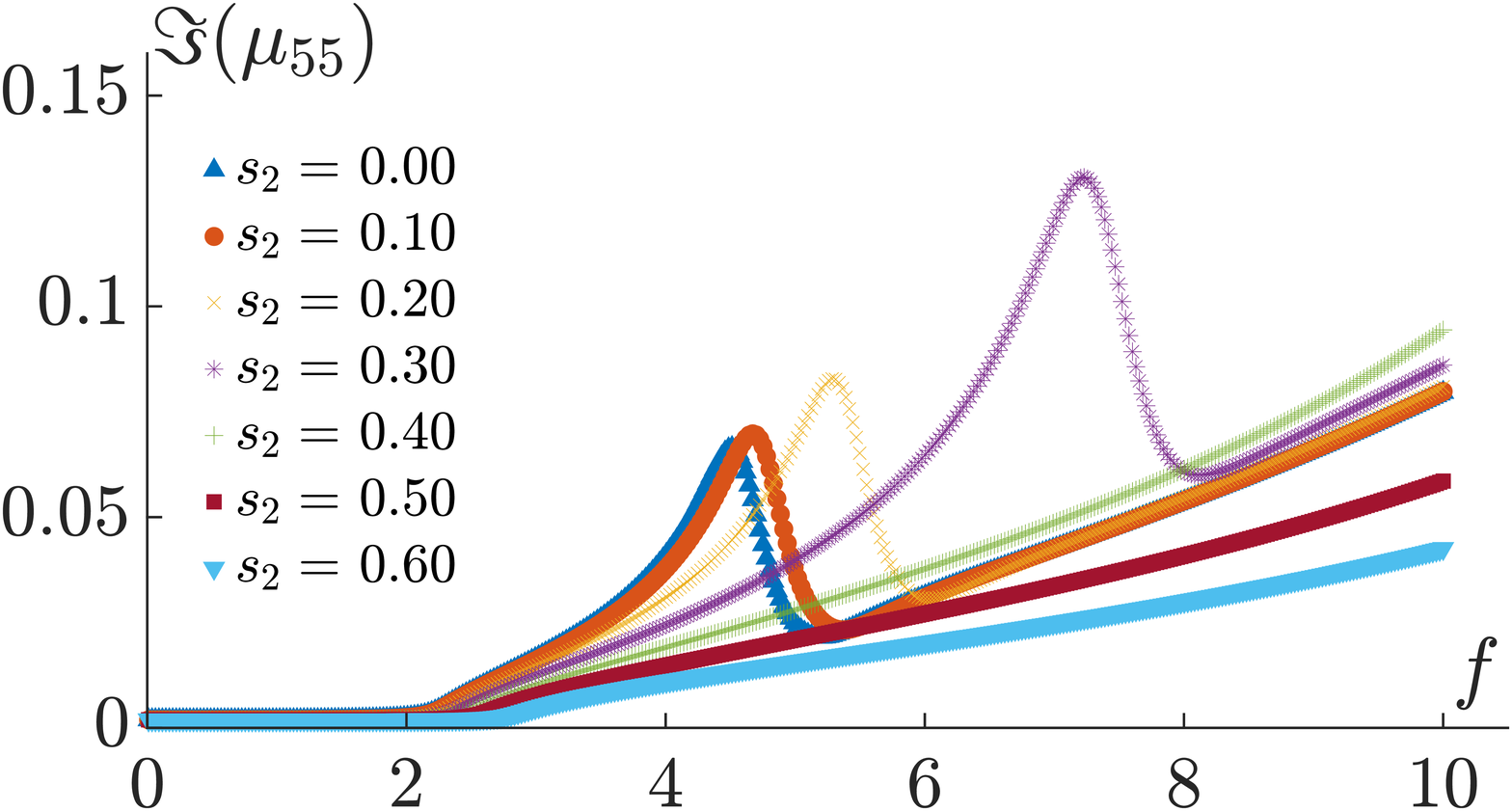}
			\caption{\label{fig:in_im_mu55_sh_sym_2}}
		\end{subfigure}
		\begin{subfigure}[b]{0.5\linewidth}
			\centering\includegraphics[height=95pt,center]{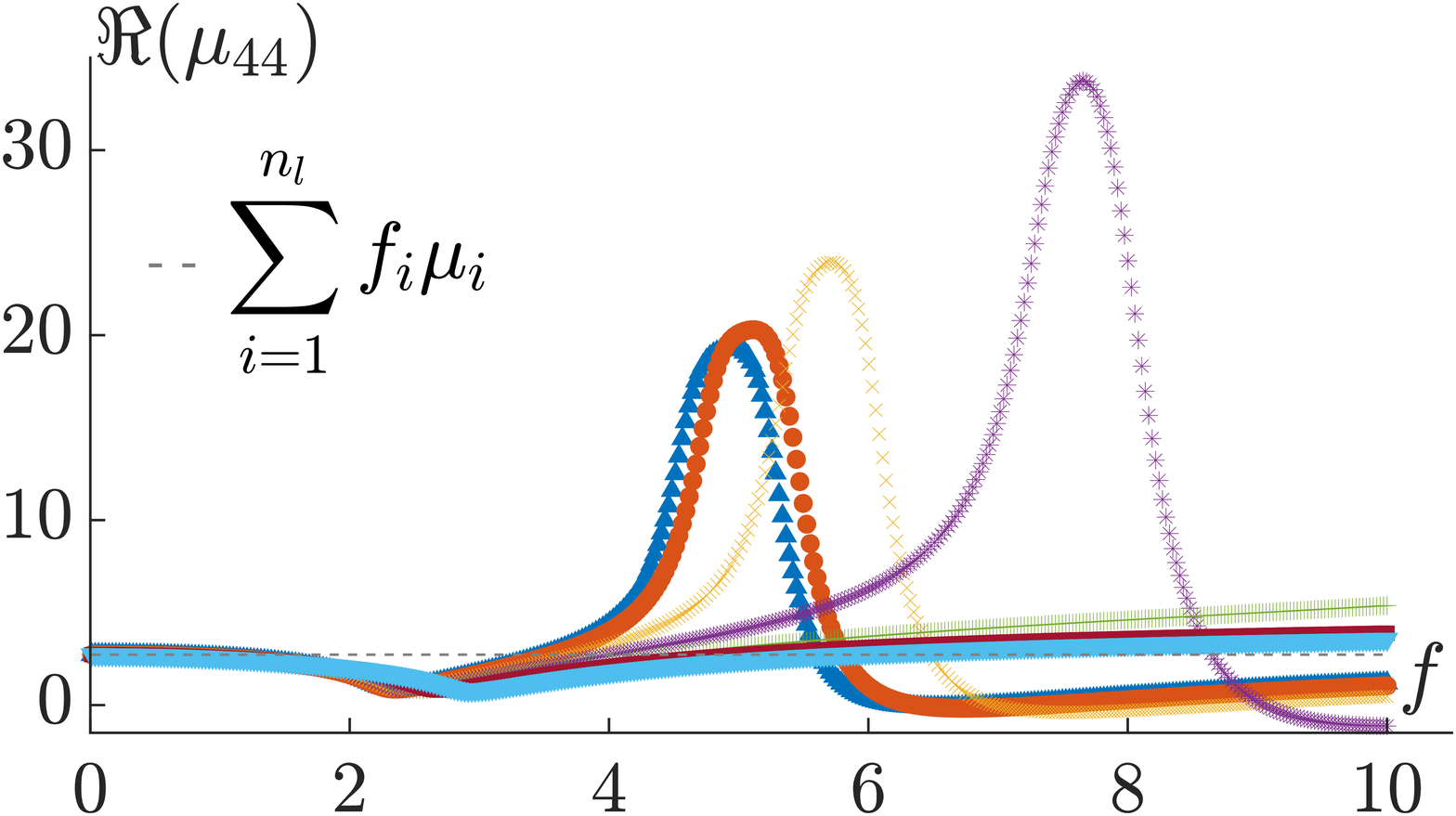}
			\caption{\label{fig:in_re_mu44_sh_sym_2}}
		\end{subfigure}%
		\begin{subfigure}[b]{0.5\linewidth}
			\centering\includegraphics[height=95pt,center]{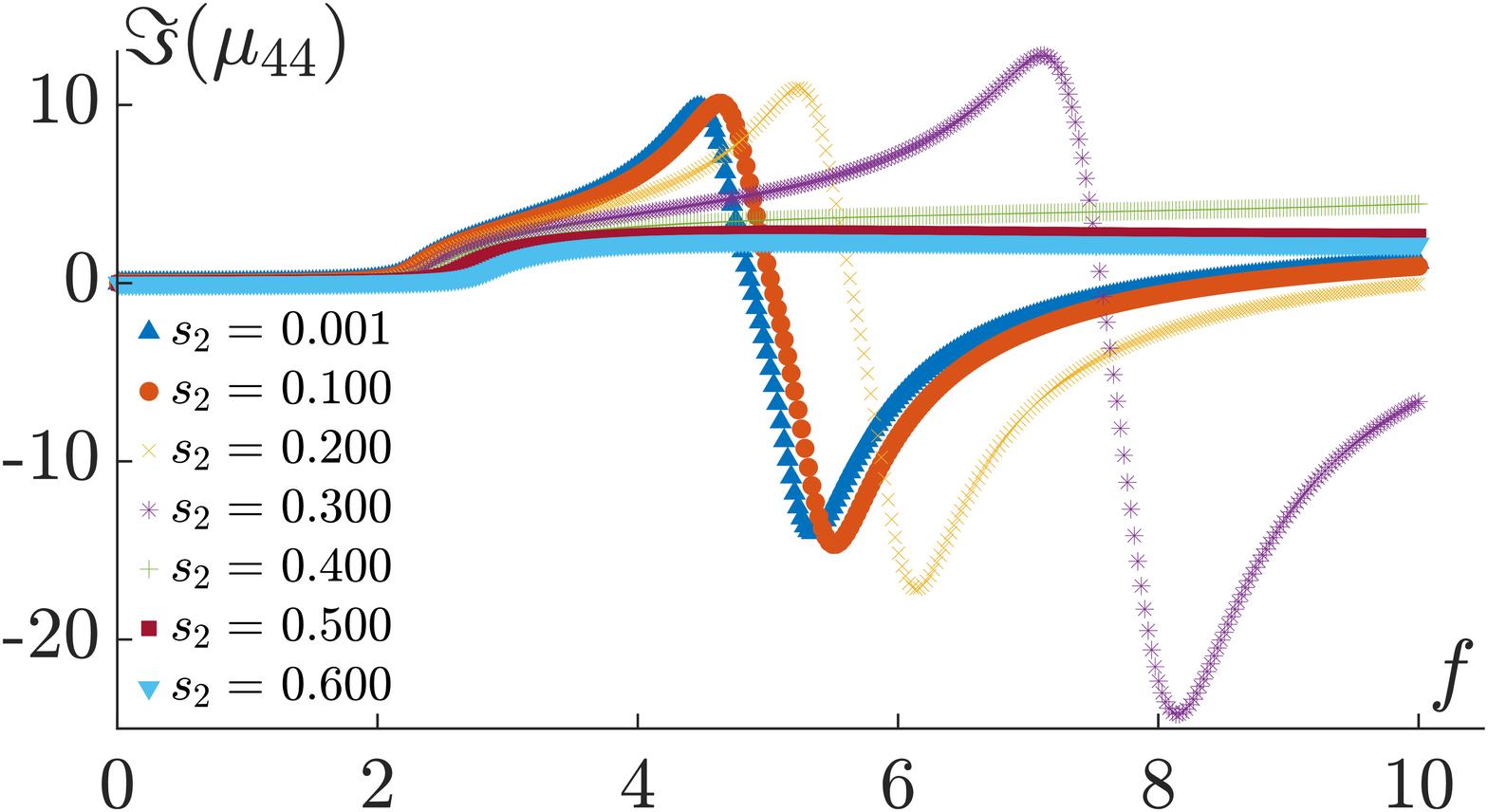}
			\caption{\label{fig:in_im_mu44_sh_sym_2}}
		\end{subfigure}
		\caption{\label{fig:sym_dispersive_diag}The only non-zero homogenized parameters, i.e. diagonal components, of the Willis constitutive tensor, for the symmetric RUC shown in Fig.~(\ref{fig:mat}). (\subref{fig:in_re_eta33_sh_sym_2}), (\subref{fig:in_re_mu55_sh_sym_2}), and (\subref{fig:in_re_mu44_sh_sym_2}) show the real components of $\eta_{33}$, $\mu_{55}$, and $\mu_{44}$, and (\subref{fig:in_im_eta33_sh_sym_2}), (\subref{fig:in_im_mu55_sh_sym_2}), and (\subref{fig:in_im_mu44_sh_sym_2}) show their imaginary counterparts, respectively. Compared to the asymmetric case shown in Fig.~(\ref{fig:asym_dispersive_diag}) the resonance peaks shifted slightly toward lower frequencies for the symmetric RUC. Note also that, as expected, the small structural difference changes the overall constitutive tensors essentially in a continuous fashion, i.e., the difference between these values and those shown in Fig.~(\ref{fig:asym_dispersive_diag}) vanish as the asymmetric unit cell converges to the symmetric one.}
	\end{figure}

	\paragraph{Overall properties based on integration} An asymmetric configuration of the RUC shown in Fig.~\ref{fig:mat} is adapted by changing the thickness and properties of layer $j = 4$ as follows: $d^4 = 3.30$, $c'^4_{T} = 0.16$, $\rho^4 = 0.38$, and $c''^4_T/c'^4_T = 0.06$. All other layers stayed intact. Figs.~(\ref{fig:asym_dispersive_diag},\ref{fig:asym_dispersive_off_diagonal}) show the real and imaginary parts of all the Willis constitutive parameters for this asymmetric unit cell. The presence of spatial dispersion can be observed in all the homogenized values. It can be seen that at the long-wavelength limit at small $s_2$ values (near normal incidence) the coupling constant $\kap_{34}$ has a linear dependence on the wave vector. More explicitly, $\kap_{34} \approx -0.019\ii k_2$, which is fully imaginary. The other coupling constitutive parameters approach 0 faster than a linear function of $k_2$.
Fig.~(\ref{fig:sym_dispersive_diag}) depicts the real and imaginary parts of the diagonal components of the constitutive tensor for the symmetric version of the RUC shown in Fig.~(\ref{fig:mat}). As mentioned earlier, for symmetric unit cells the off-diagonal terms become identically zero. All three diagonal terms stay spatially dispersive.
	
\begin{figure}[!ht]
		\begin{subfigure}[b]{0.5\linewidth}
			\centering\includegraphics[height=95pt,center]{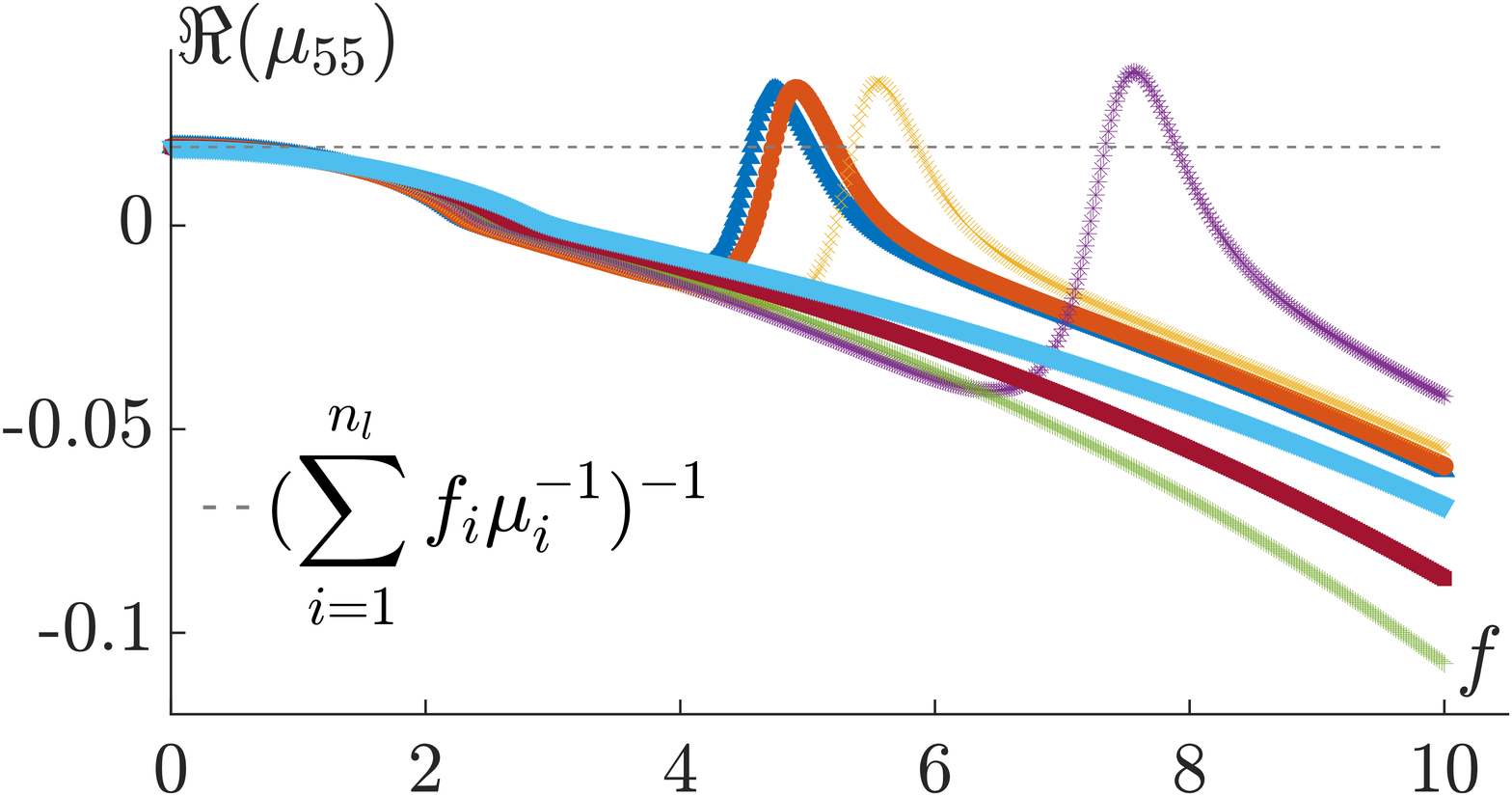}
			\caption{\label{fig:os_re_mu55_sh}}
		\end{subfigure}%
		\begin{subfigure}[b]{0.5\linewidth}
			\centering\includegraphics[height=95pt,center]{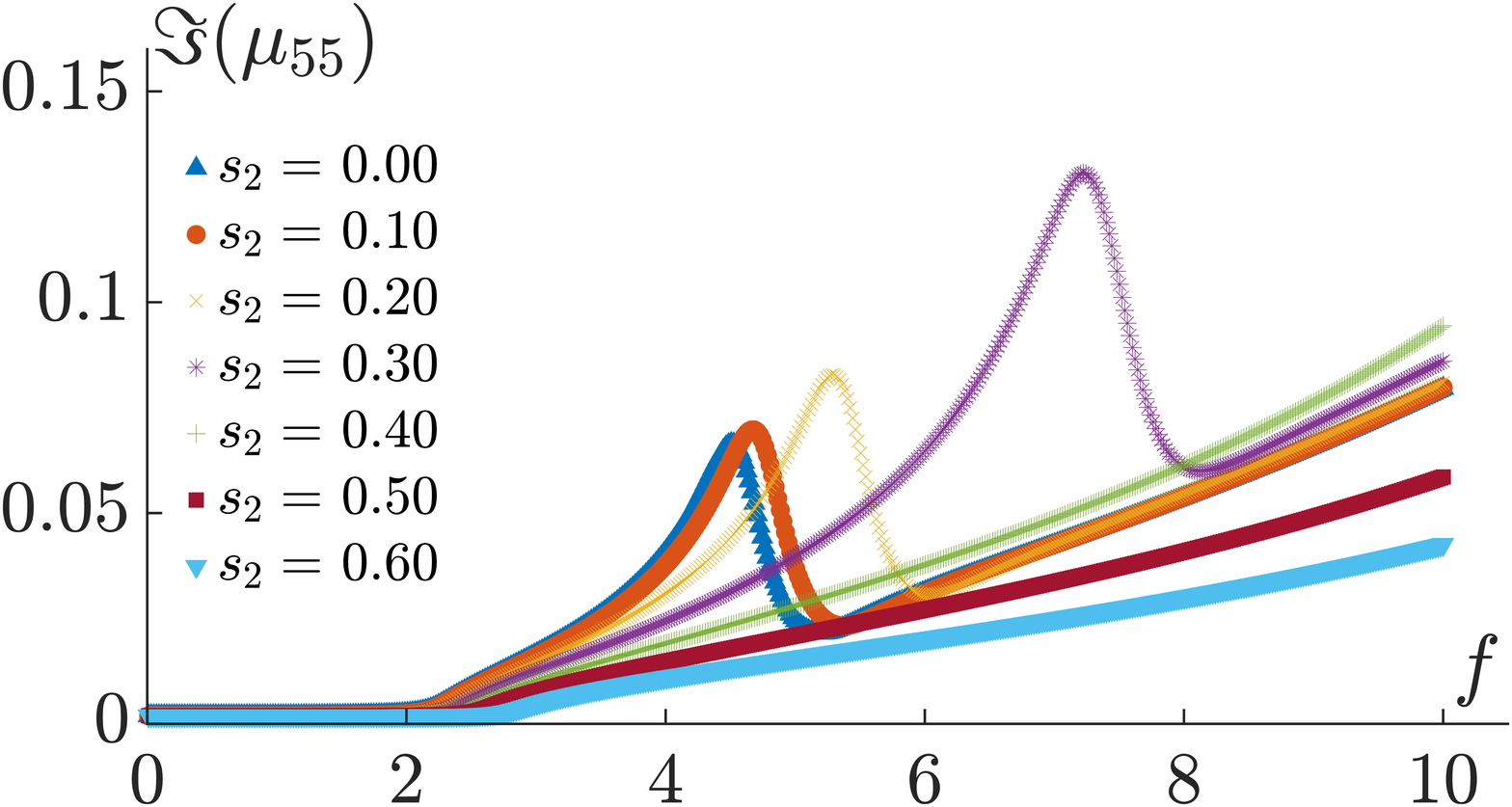}
			\caption{\label{fig:os_im_mu55_sh}}
		\end{subfigure}
		\begin{subfigure}[b]{0.5\linewidth}
			\centering\includegraphics[height=95pt,center]{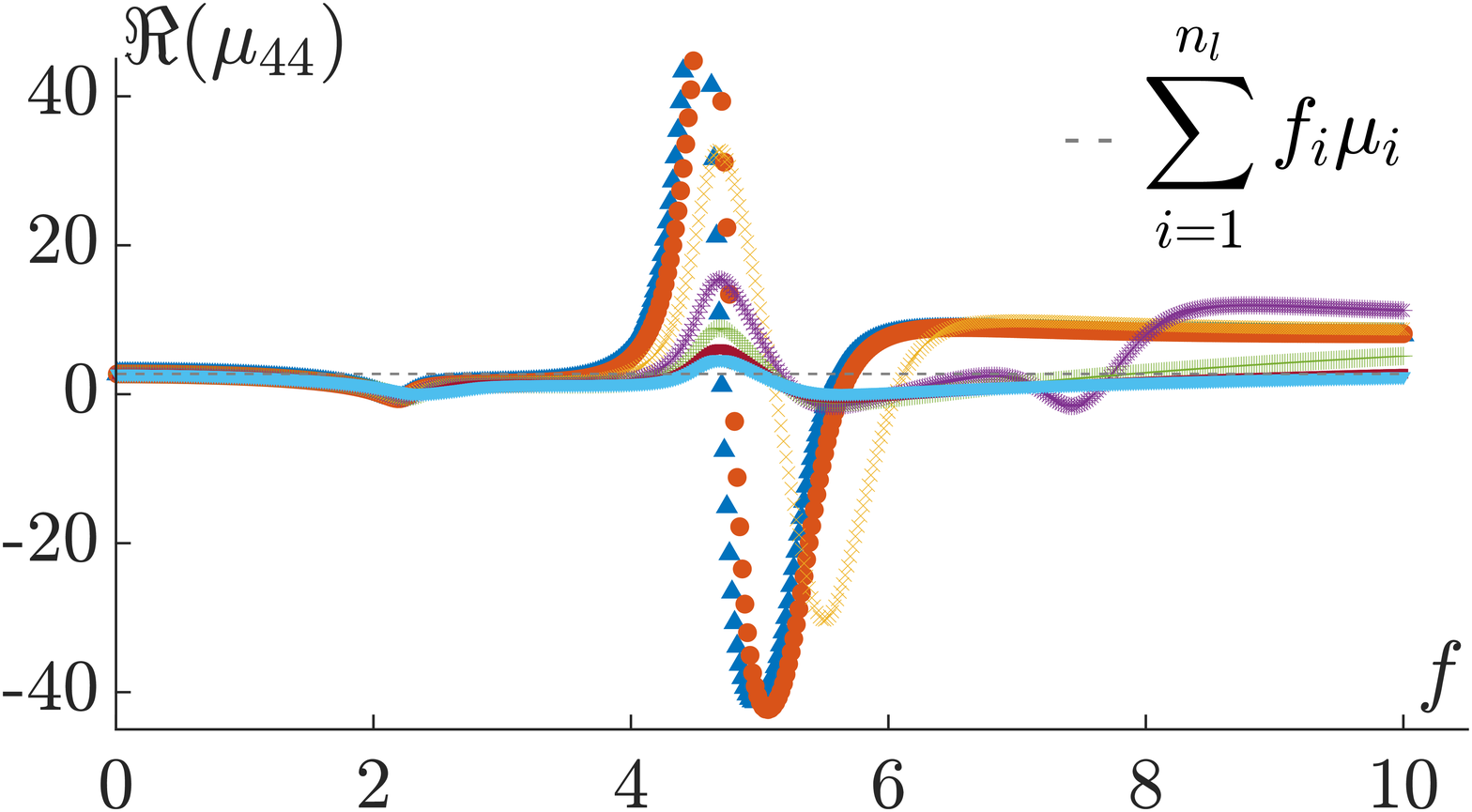}
			\caption{\label{fig:os_re_mu44_sh}}
		\end{subfigure}%
		\begin{subfigure}[b]{0.5\linewidth}
			\centering\includegraphics[height=95pt,center]{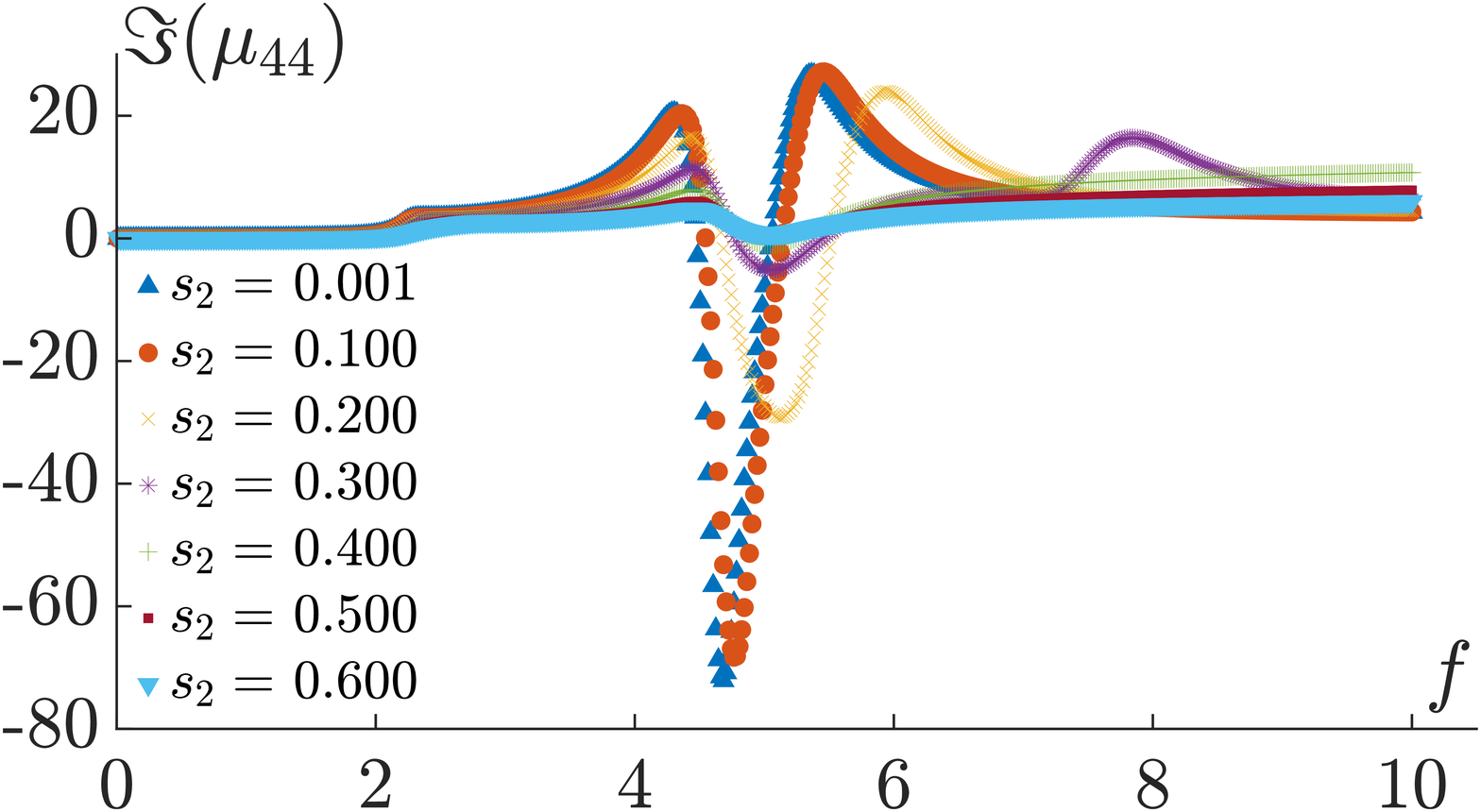}
			\caption{\label{fig:os_im_mu44_sh}}
		\end{subfigure}
		\begin{subfigure}[b]{0.5\linewidth}
			\centering\includegraphics[height=95pt,center]{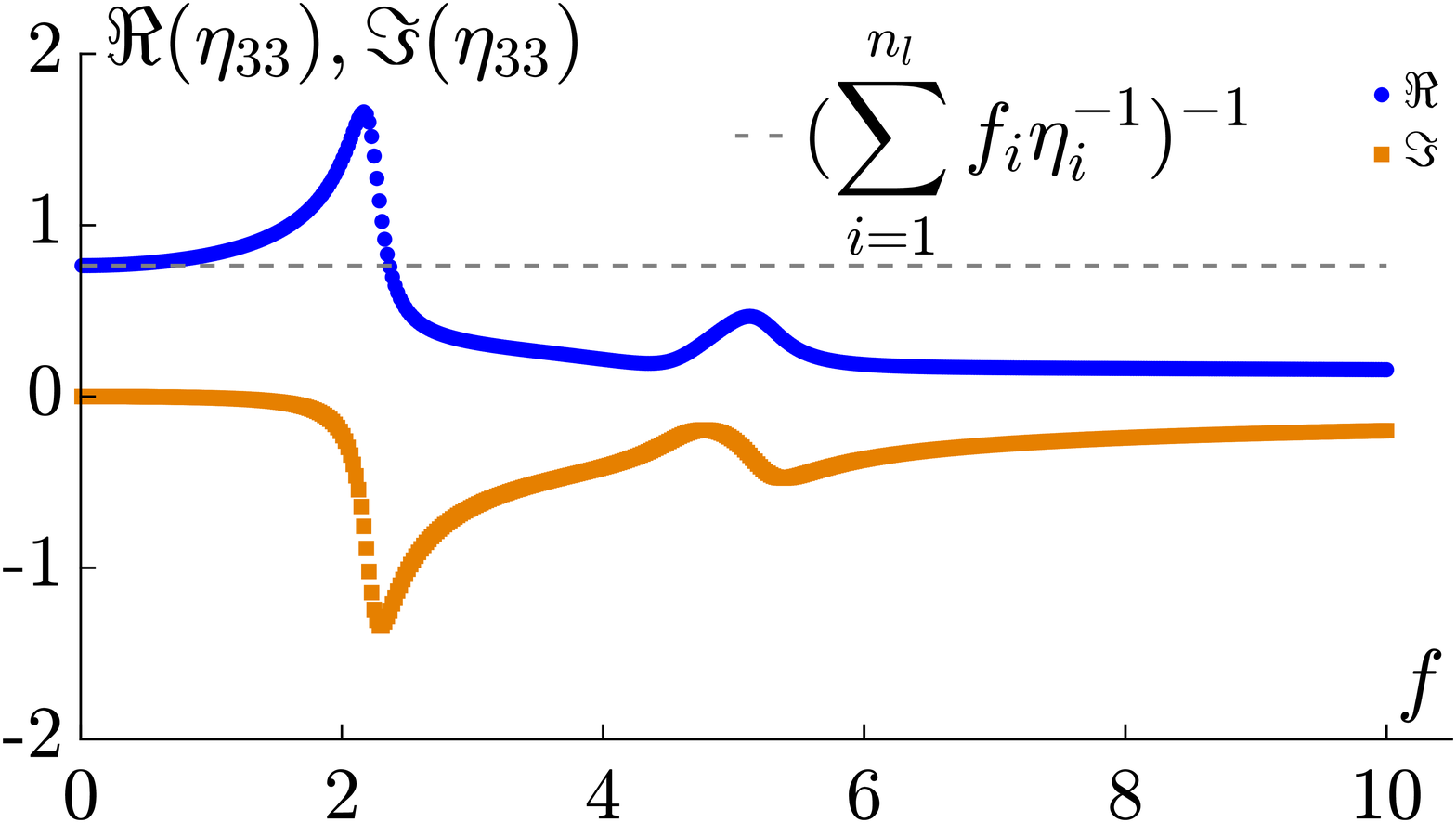}
			\caption{\label{fig:os_eta33_sh}}
		\end{subfigure}%
		\begin{subfigure}[b]{0.5\linewidth}
			\centering\includegraphics[height=95pt,center]{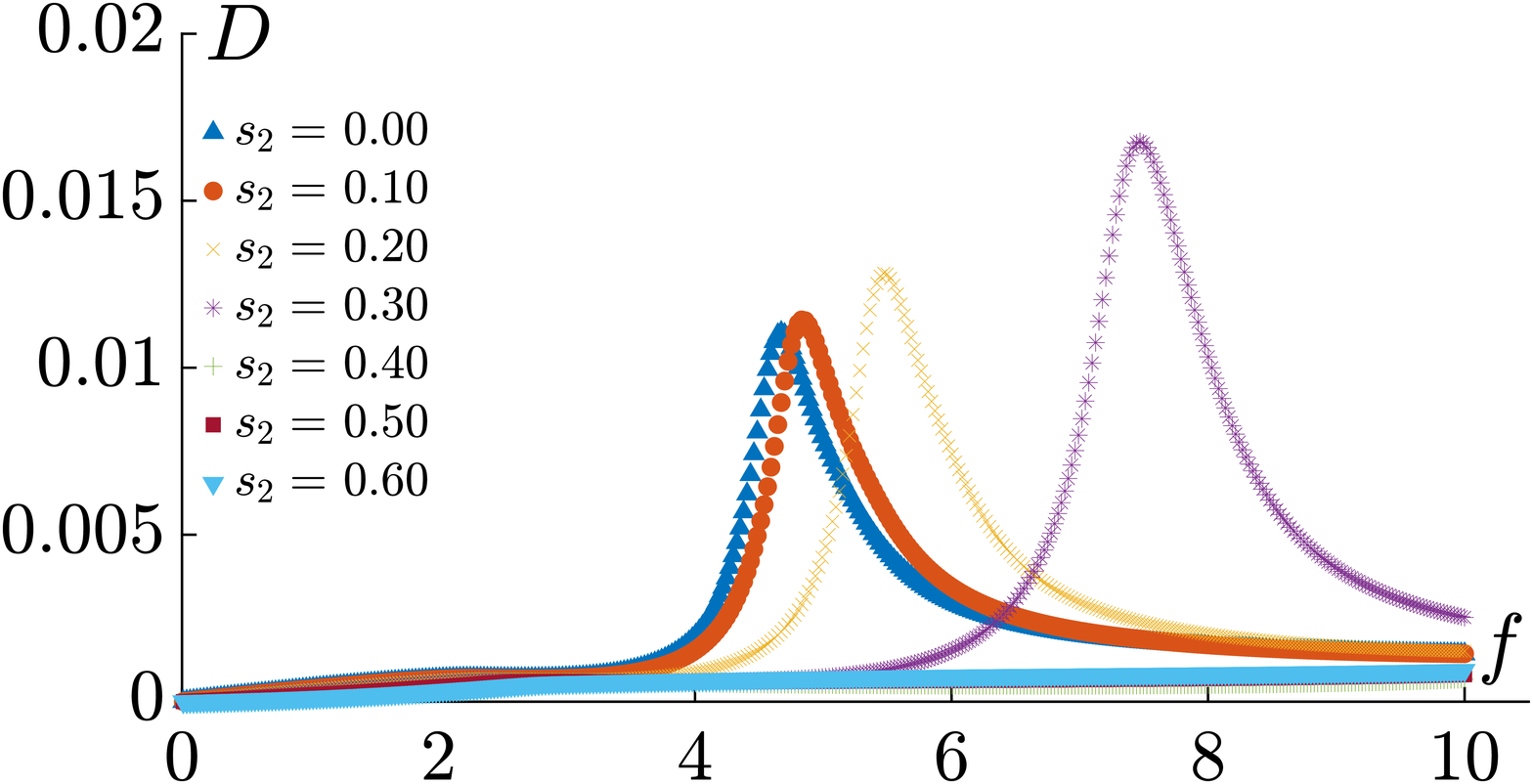}
			\caption{\label{fig:pass}}
		\end{subfigure}
		\caption{\label{fig:os_sc1} Apparent constitutive functions based only on scattering for a fully diagonal constitutive tensor with spatially non-dispersive $\eta_{33}$. Real and imaginary parts of $\mu_{55}$ and $\mu_{44}$ are shown in (\subref{fig:os_re_mu55_sh}), (\subref{fig:os_im_mu55_sh}) and (\subref{fig:os_re_mu44_sh}), (\subref{fig:os_im_mu44_sh}). At low frequencies, $\eta_{33}$ (\subref{fig:os_eta33_sh}) and $\mu_{55}$ match their associated Reuss models (iso-stress or volume average for inverse stiffness and volume average for density, i.e., inverse $\eta_{33}$) and $\mu_{44}$ also approaches its Voigt model (iso-strain or volume average for stiffness), which are all expected at the long-wavelength approximation. Unlike the 3-layer structure yielding positive $\mu_{55}$ over the entire frequency-slowness plane as shown in \cite{Amirkhizi2018}, negative real parts of $\mu_{55}$ are observed starting from around \SI{2.5}{kHz}. (\subref{fig:pass}) The positive time-averaged dissipation in an RUC shown in Fig.~(\ref{fig:mat}) for oblique anti-plane shear waves at any frequency, based on Eq.~\ref{eq:pass_sym}. For elastic systems $D = 0$ (not shown here).}
	\end{figure}

	\paragraph{Comparison with the overall properties derived based only on scattering} From an experimental perspective the details of the fields inside a unit cell are not always available and the interaction of the sample with exterior domain is entirely determined based on its scattering or more precisely the field components that are continuous across boundaries. With such limited information, the overall constitutive tensors may not be uniquely determined. Different sets of assumptions may be enforced to derive overall constitutive tensors that will reproduce the same scattering of the heterogeneous slab from a homogenized slab of the same thickness \cite{Amirkhizi2018}. The scattering data does not even uniquely determine a presumed diagonal constitutive tensor for a symmetric RUC. A unique candidate may be determined from scattering if one assumes that $\eta_{33}$ is spatially non-dispersive. In other words, the value determined for normal incidence ($s_2 = 0$) is used for all other wave vectors. The other two diagonal components ($\mu_{44}, \mu_{55}$) will still need to be spatially dispersive. Figure~(\ref{fig:os_sc1}) shows the overall properties based on such an approach. When compared to the integration results, Fig.~(\ref{fig:sym_dispersive_diag}), $\mu_{55}$ matches between both methods for all values of $s_2$, as do the values of all quantities, at $s_2=0$. While neither of these two sets contradict Onsager or Betti-Maxwell reciprocity, passivity (see below), or causality, $\mu_{44}$ of the two approaches are not similar. The difference is a mere consequence of enforcing locality for $\eta_{33}$.
	
	We must note that in \cite{Amirkhizi2018} a different case was studied, in which the diagonal terms were enforced to be spatially non-dispersive. Furthermore, $\kap_{35}$ may be considered a free parameter and was set it to zero for a symmetric cell considering the limiting case of normal incidence. To maintain reciprocity in the  dispersion equation, Eqs.~\eqref{eq:disp}, leads to $\kap_{34} = -\kap_{43}$, and $\kap_{53} = - \kap_{35} = 0$, which will then also require $\mu_{54} = - \mu_{45}$. Such assumption although contradicts the strong from of Betti-Maxwell reciprocity relation if $\mu_{45} \neq 0$ (see Eq.~\eqref{eq:mu_sym}), can still reproduce scattering, precisely. Note again that this strong form is only a sufficient condition, but not necessary, for reciprocity of the system. 
	\paragraph{Note on passivity} If there are no energy sources within a domain $ \mathcal{V}$, the integration of energy flux into its closed boundary, $\partial\mathcal{V}$, should yield zero or positive values, for lossless and lossy passive media \cite{Srivastava2015, Amirkhizi2017,srivastava2020causality}:
	\begin{equation*}
		\begin{aligned}
			\Psi &= \int_{\partial \vol}t_jv_j \ \dd\area = \int_{\partial\vol} n_j\sigma_{ij}v_j \ \dd\area = \int_{\vol} (\sigma_{ij} v_j)_{,i} \ \dd\vol=\int_{\vol}(\sigma_{ij,i}v_j + \sigma_{ij}v_{j,i}) \ \dd\vol\\
			&= \int_{\vol}p_{j,t}v_j + \sigma_{ij}\left(\varepsilon_{ij,t} + w_{ij,t}\right) \ \dd\vol = \int_{\mathcal{V}} (p_{j,t}v_j + \sigma_{ij}\varepsilon_{ij,t}) \ \dd\vol.
		\end{aligned}
	\end{equation*}
	Considering $\ee^{\ii(\omega t - \vect{k.x})}$ phasor convention,
	\begin{equation*}
		\Psi = \int_{\mathcal{V}} \Re(\ii\omega p_{j})\Re(v_j) + \Re(\sigma_{ij})\Re(\ii\omega\varepsilon_{ij}) \ \dd\mathcal{V}.
	\end{equation*}
	Time averaging of the real values of $\omega$ gives:
	\begin{equation*}	
		\dfrac{1}{2\pi/\omega}\int_{t}^{t+2\pi/\omega} \Psi \ \dd t = \dfrac{\omega}{2}\int_{\mathcal{V}}\Re(\ii p_jv_j^* + \ii\sigma_{ij}^*\varepsilon_{ij}) \ \dd\mathcal{V}= \dfrac{\omega}{2}\int_{\mathcal{V}}-\Im(p_jv_j^* + \sigma_{ij}^*\varepsilon_{ij}) \ \dd\vol \geqslant 0,
	\end{equation*}
	where $^*$ denotes complex conjugate. For the case of propagating oblique SH-waves and positive $\omega$, using Voigt's notation:
	\begin{equation}
		D = -\dfrac{\omega}{2}\Im(T_5^*\Gamma_5 + T_4^*\Gamma_4 + P_3V_3^*)\geqslant 0,
	\end{equation}
	in which the equality holds for lossless cases. This can be further simplified to:
	\begin{equation}\label{eq:passivity}
		D = -\omega \left(s''_2 Z'_{43}+S''_1 Z'_{53}\right) \geqslant 0,
	\end{equation}
	where superscripts $'$ and $''$ refer to the real and imaginary parts, and ${S_1(\omega, s_2) = K_1/\omega}$. In general, the impedances for general Willis media (see \cite{Amirkhizi2018}) 
	\begin{equation}
		\begin{aligned}
			Z_{43}(\omega, s_2) &= -\dfrac{T_4}{V_3} = \mu_{44} s_2 + \mu_{45} S_1 -(\kap_{43}/\eta_{33}) (1 + \kappa_{34} s_2 + \kappa_{35} S_1),\\
			Z_{53}(\omega, s_2) &= -\dfrac{T_5}{V_3} =\mu_{55} S_1 + \mu_{54} s_2-(\kappa_{53}/\eta_{33}) (1 + \kappa_{35} S_1 + \kappa_{34} s_2),
		\end{aligned}
	\end{equation}
	can be used along with the dispersion equation to convert this into a single equation in terms of constitutive parameters for each value of $s_2$.
	
	The general form is too complicated to be of any conceptual use. However, for symmetric RUCs with zero off-diagonal constitutive parameters and using Eq.~\eqref{eq:const} and the simplified dispersion relation, the passivity condition would simplify to,
	\begin{equation}\label{eq:pass_sym}
		\dfrac{\eta_{33}''}{|\eta_{33}|} + \dfrac{\mu_{44}''}{|\mu_{44}|}\left|\eta_{33}\mu_{44}s_2^2\right| + \dfrac{\mu_{55}''}{|\mu_{55}|}\left|\eta_{33}\mu_{55}S_1^2\right| \geqslant 0,
	\end{equation}
	where $\eta_{33}\mu_{55}S_1^2 =  1 - \eta_{33}\mu_{44}s_2^2$ can be used to eliminate $S_1$, and equality holds for lossless systems. 
	
	All the cases and results presented in this work satisfy the passivity condition, with equality to zero for the lossless cases. An example can be seen in Figure~(\ref{fig:pass}) where the value of dissipation time average, $D$, is plotted as a function of frequency for different $s_2$ values for the symmetric and lossy RUC shown in Fig.~(\ref{fig:mat}). The fully elastic systems with symmetric RUC, despite of exhibiting a non-Hermitian constitutive tensor (complex diagonal parameters) the passivity requirement shown in Eq.~\eqref{eq:passivity} is satisfied at all frequencies and $s_2$ values, indicating that the much stronger Hermiticity condition for the constitutive tensor of fully elastic systems is enough but not necessary. The main reason, as was previously mentioned in \cite{Amirkhizi2017}, is that the particle velocity and stress or particle velocity and momentum are not linearly independent for waves that exist in the absence of body sources. While it is possible to create waves with independent particle velocities and stress values by introducing specific profiles of body forces, the energy balance needs to consider the work of such sources, which will interfere with the calculation of energy loss due to material dissipation and therefore is not considered here. Furthermore, introduction of body sources may be utilized to create waves away from dispersion surfaces, which is beyond the scope of the present work. 
	
	\section{Summary and conclusions}
	The overall constitutive parameters of layered media are naturally defined as those that will produce the same scattering response as the original heterogeneous specimen for a presumably homogeneous sample with the same overall geometry (e.g., a slab of the same thickness). For the oblique incidence of anti-plane shear waves, this requirement does not uniquely determine the entirety of the Wills-type constitutive matrix. A consistent field averaging technique is proposed here which is naturally compatible with the dispersion and scattering response. Using this integration scheme, a systematic method to calculate all the Willis parameters in layered media is presented. To calculate unique constitutive parameters, Onsager's principle and reciprocity are used. As previously established, spatially dispersive constitutive parameters are required to fully match the scattering of the heterogeneous media. While Betti-Maxwell's reciprocity is not a necessary condition for spatially dispersive media, they are used in this work to lead to a unique choice of constitutive tensor. It is observed that for a symmetric unit cell, a fully diagonal constitutive matrix is derived, however, all the non-zero terms are functions of the wave vector and have resonance peaks at different frequencies for each propagation direction. In the abscess of the $x_1$-symmetry, i.e., asymmetric cells, non-zero spatially dispersive off-diagonal terms are required in overall description of the system. All the parameters approach their appropriate Voigt or Reuss averages as $\omega \rightarrow 0$. However, at the long-wavelength limit and even for near-normal incidence, the nonlocality does not vanish as $\kap_{34}$ can be approximated by a fully imaginary linear function of $k_2$.  Finally, the passivity of such systems is studied, and it is established that for source-free waves, the derived constitutive tensors satisfy the necessary condition for the time-average of energy loss to be positive definite, along with its vanishing for fully elastic systems. It is worth mentioning that the similar process may be applied to more complex 3D micro-structures. The number of unknown constitutive parameters will increase, as do the number of equations associated with Onsager principle and Betti-Maxwell reciprocity. While such generalization will be addressed in a future publication, it appears that a reformulation based on the full wave vector as the independent parameter may be required. Therefore, the current solution for wave vector component as a function of frequency will have to be modified.
	
	\section*{Acknowledgement}
	This work has been performed with partial support from the University of Massachusetts and National Science Foundation (NSF) grant \#1825969 to the University of Massachusetts, Lowell.
	
	\appendix
	\section{Transfer matrix method}
	\label{app:tm}
	The transfer matrix for each layer is the linear operator that relates the values of the state vector $\vect{\psi}$ on the two faces of that layer and can be written as,
	\begin{equation}
		\vect{\TM}^j(k_2) = \vect{\zeta}^j\vect{\delta}^j(d^j,k_2)({\vect{\zeta}^j})^{-1}.
	\end{equation}
	Based on the overall continuity of the $\vect{\psi}$, the transfer matrix of any multilayer medium can then be easily calculated as,
	\begin{equation}
		\vect{\TM} = \vect{\TM}^{N_l}\vect{\TM}^{N_{l-1}}\cdots\vect{\TM}^j\cdots\vect{\TM}^1.
	\end{equation}
	For an infinitely periodic system generated from a repeating unit cell, the values of $\vect{\psi}$ evaluated at the boundaries of the RUC must conform to Bloch-Floquet periodicity, i.e.
	\begin{equation} \label{eq:tm-eigen}
		\vect{\TM} \vect{\psi}(x_1 = x_1^0) = \vect{\psi}(x_1 = x_1^0 + d) = \ee^{-\ii Q^\alpha} \vect{\psi}(x_1 = x_1^0).
	\end{equation}
	Furthermore, the apparent impedance of the cell can be extracted from the eigenvectors 
	\begin{align}
		Q^\alpha &=  \ii\log\left[\dfrac{1}{2}\left(\TM_{11} + \TM_{22}\mp\sqrt{(\TM_{11}-\TM_{22})^2 + 4\TM_{12}\TM_{21}}\right)\right], \label{eq:tm-Q}\\
		Z_{53}^\alpha &= -\dfrac{T_5^\alpha}{V_3^\alpha} = \dfrac{2\TM_{21}}{\TM_{22}-\TM_{11}\pm\sqrt{(\TM_{11}-\TM_{22})^2 + 4\TM_{12}\TM_{21}}}.\label{eq:tm-Z}
	\end{align} 
	One can use these to define an overall wave vector component $K_1^\alpha = Q^\alpha / d$. The superscript $\alpha = \pm$ identifies the two possible solutions traveling towards positive and negative $x_1$ values. Following \cite{Amirkhizi2017}, $2 n \pi$ may be added or subtracted from $Q$ whenever needed to achieve continuity (if possible) and remove the ambiguity in the phase. The continuity of the phase is not guaranteed in 2D systems; see \cite{Abedi2020}.
	
	\section{Scattering analysis}
	\label{app:scatt}
	For a finite layered slab between two semi-infinite homogeneous domains, the complex wave amplitudes of the right- and left-hand side of the layered medium can be related as well 
	\begin{equation}
		\vect{A}^b = \vect{\PM}\vect{A}^a.
	\end{equation}
	Here the superscripts $^b$ and $^a$ identify both the semi-infinite domains as well as the complex wave amplitudes in them measured at locations $x_1 = x^{a/b}$. $\PM$ is also frequently called the transfer matrix in literature e.g., in \cite{markos2008wave} and can be obtained from the transfer matrix as 
	\begin{equation} \label{eq:pmat-tmat}
		\vect{\PM} = (\vect{\delta}^b(x_1=x^b,k_2))^{-1}(\vect{\zeta}^b)^{-1}\vect{\TM}\vect{\zeta}^a\vect{\delta}^a(x_1=x^a,k_2).
	\end{equation}
	The new polarization matrices are exactly similar to before, and the phase advance matrices can be written as
	\begin{equation}
		\begin{aligned}
			\vect{\delta}^a(x_1,k_2) &=
			\begin{pmatrix}
				\ee^{-\ii k_1^{a+}(x_1 - x^{l})} & 0\\
				0 & \ee^{-\ii k_1^{a-}(x_1 - x^{l})}
			\end{pmatrix}, \\
			\vect{\delta}^b(x_1,k_2) &=
			\begin{pmatrix}
				\ee^{-\ii k_1^{b+}(x_1 - x^{r})} & 0\\
				0 & \ee^{-\ii k_1^{b-}(x_1 - x^{r})}
			\end{pmatrix}, 
		\end{aligned}
	\end{equation}
	where $x_1 = x^{l/r}$ represent the left and right faces of the sample. With simplification $x^{a/b} = x^{l/r}$, the two phase advance matrices will become identity. The components of the scattering matrix,
	\begin{equation}
		\begin{pmatrix}
			A^{a-}\\
			A^{b+}
		\end{pmatrix}=
		\begin{pmatrix}
			\SM_{aa} & \SM_{ab}\\
			\SM_{ba} & \SM_{bb}
		\end{pmatrix}
		\begin{pmatrix}
			A^{a+}\\
			A^{b-}
		\end{pmatrix},
	\end{equation}
	which relates outgoing to incoming (incident) waves, can be written as 
	\begin{equation}
		\begin{aligned} 
			\SM_{aa} &= -\PM_{21}/\PM_{22},\\
			\SM_{ba} &= \PM_{11} - \PM_{12}\PM_{21}/\PM_{22},\\
			\SM_{bb} &= \PM_{12}/\PM_{22},\\
			\SM_{ab} &= 1/\PM_{22}.
		\end{aligned}
	\end{equation}
	One can confirm that when $\det(\TM)=1$, as a result of reciprocity, and with $Z^a = Z^b$, then $\SM_{ab} = \SM_{ba}$ \cite{wang2020exceptional}.
	
	\bibliographystyle{unsrtm}
	\bibliography{SH3}
	
\end{document}